\documentclass[structabstract]{aa}  
\usepackage{graphicx}
\usepackage{longtable}
\usepackage{lscape}
\usepackage{amssymb}
\usepackage{endnotes} 
\usepackage{subfigure}
\usepackage{txfonts}
\usepackage{natbib}
\usepackage{url}
\usepackage{color}  
\begin{document}

\title{Investigation of the stellar content in the western part of the Carina nebula
\thanks{Table 3 is only available in electronic form
at the CDS via anonymous ftp to cdsarc.u-strasbg.fr (130.79.128.5)
or via http://cdsweb.u-strasbg.fr/cgi-bin/qcat?J/A+A/}}

\author{Brajesh Kumar\inst{1,2} 
\and Saurabh Sharma\inst{2} 
\and Jean Manfroid\inst{1}
\and Eric Gosset\inst{1}\fnmsep\thanks{Senior Research Associate FNRS}
\and \\ Gregor Rauw\inst{1} 
\and Ya\"{e}l Naz{\'e}\inst{1}\fnmsep\thanks{Research Associate FNRS}
\and Ram Kesh Yadav\inst{2} }

\institute{Institut d'Astrophysique et de G\'{e}ophysique, Universit\'{e} de
Li\`{e}ge, All\'{e}e du 6 Ao\^{u}t 17, B\^{a}t B5c, 4000 Li\`{e}ge, Belgium \\
\email{brajesh@aries.res.in, brajesharies@gmail.com} 
\and Aryabhatta Research Institute of Observational Sciences, Manora Peak,
Nainital 263 002, India}

\date{Received --- / Accepted ---}
\abstract
{The low obscuration and proximity of the Carina nebula make it an ideal place to study the ongoing
star formation process and impact of massive stars on low-mass stars in their surroundings.}
{To investigate this process, we generated a new catalog of the pre-main-sequence (PMS) 
stars in the Carina west (CrW) region and studied their nature and spatial distribution. We  
also determined various parameters (reddening, reddening law, age, mass), which are used further 
to estimate the initial mass function (IMF) and $K$-band luminosity function (KLF) for the 
region under study.}
{We obtained deep $UBVRI$ $H\alpha$ photometric data of the field situated to the west of the main
Carina nebula and centered on WR~22. Medium-resolution optical spectroscopy of a subsample of
X-ray selected objects along with archival data sets from $Chandra$, $XMM-Newton$ and 2MASS surveys 
were used for the present study. Different sets of color-color and color-magnitude diagrams are
used to determine reddening for the region and to identify young stellar objects (YSOs) and estimate
their age and mass.}
{Our spectroscopic results indicate that the majority of the X-ray sources are late spectral type 
stars. The region shows a large amount of differential reddening with minimum and maximum values of 
$E(B-V)$ as 0.25 and 1.1 mag, respectively. Our analysis reveals that the total-to-selective absorption 
ratio $R_V$ is $\sim$3.7 $\pm$ 0.1, suggesting an abnormal grain size in the observed region. 
We identified 467 YSOs and studied their characteristics. The ages and 
masses of the 241 optically identified YSOs range from $\sim$0.1 to 10 Myr and $\sim$0.3 to 
4.8 M$_\odot$, respectively. However, the majority of them are younger than 1 Myr and have masses 
below 2 M$_\odot$. 
The high mass star WR~22 does not seem to have contributed to the formation of YSOs in the CrW 
region. The initial mass function slope, $\Gamma$, in this region is found to be $-$1.13 $\pm$ 
0.20 in the mass range of 0.5 $<$ M/M$_\odot$ $<$ 4.8. The $K$-band luminosity function slope ($\alpha$) 
is also estimated as 0.31 $\pm$ 0.01. We also performed minimum spanning tree analysis
of the YSOs in this region, which reveals that there are at least ten YSO cores associated with the 
molecular cloud, and that leads to an average core radius of 0.43 pc and a median branch length 
of 0.28 pc.}
{}

\titlerunning{Stellar content in the Carina west region}
\authorrunning{B. Kumar et al.}

\maketitle
\section{Introduction}
Massive stars (M $>$~8$-$10 M$_{\odot}$) in star-forming regions significantly influence 
their surroundings. In the course of their life, the feedback provided by their energetic 
ionization radiation and powerful stellar winds regulate the formation of low- and 
intermediate-mass stars \citep{1999PASP..111.1049G, 2007ARA&A..45..481Z}. After a 
short life time ($\lesssim$10$^{7}$ years), they explode as supernovae or hypernovae 
(supernovae with substantially higher energy than standard supernovae) enriching the 
interstellar medium with the products of the various nucleosynthesis processes that have 
occurred during their lifetime \citep[see][and references therein]{1995ARA&A..33..115A, 
1996snai.book.....A, 1995ApJS..101..181W, 2003IAUS..212..395N}. The shock waves produced 
in these events may trigger new star formation \citep[e.g.][]{1998ASPC..148..150E}. 
Characterizing the young stellar objects (YSOs) in massive star-forming regions is therefore 
of utmost importance to understand the link with the neighboring massive star population.

The Carina nebula (NGC~3372) region, which hosts several young star clusters made of very 
massive stars along with YSOs, provides an ideal laboratory for studying the ongoing star 
formation \citep[see][]{2008hsf2.book..138S}. The CO survey of this region demonstrates that 
the Carina nebula is on the edge of a giant molecular cloud extending over $\sim$130 pc and 
has a mass in excess of 5 $\times$ 10$^5$ M$_{\odot}$ \citep[see][]{1988ApJ...331..181G}. 
It contains $\sim$200 OB stars \citep{2006MNRAS.367..763S, 2011ApJS..194...14P}, more than 
$\sim$60 massive O stars \citep[see][]{1995RMxAC...2...57F, 2006MNRAS.367..763S}, and three 
WN(H)\footnote{These are late type WN stars with hydrogen; for a review of WR stars, 
see \citet{1987ARA&A..25..113A} and \citet{2007ARA&A..45..177C}.} stars (i.e. WR~22, 24, and 25; 
\citealt{2006MNRAS.367..763S, 2008hsf2.book..138S}).

\begin{figure*}
\centering\includegraphics[height=17cm,width=17cm,angle=0]{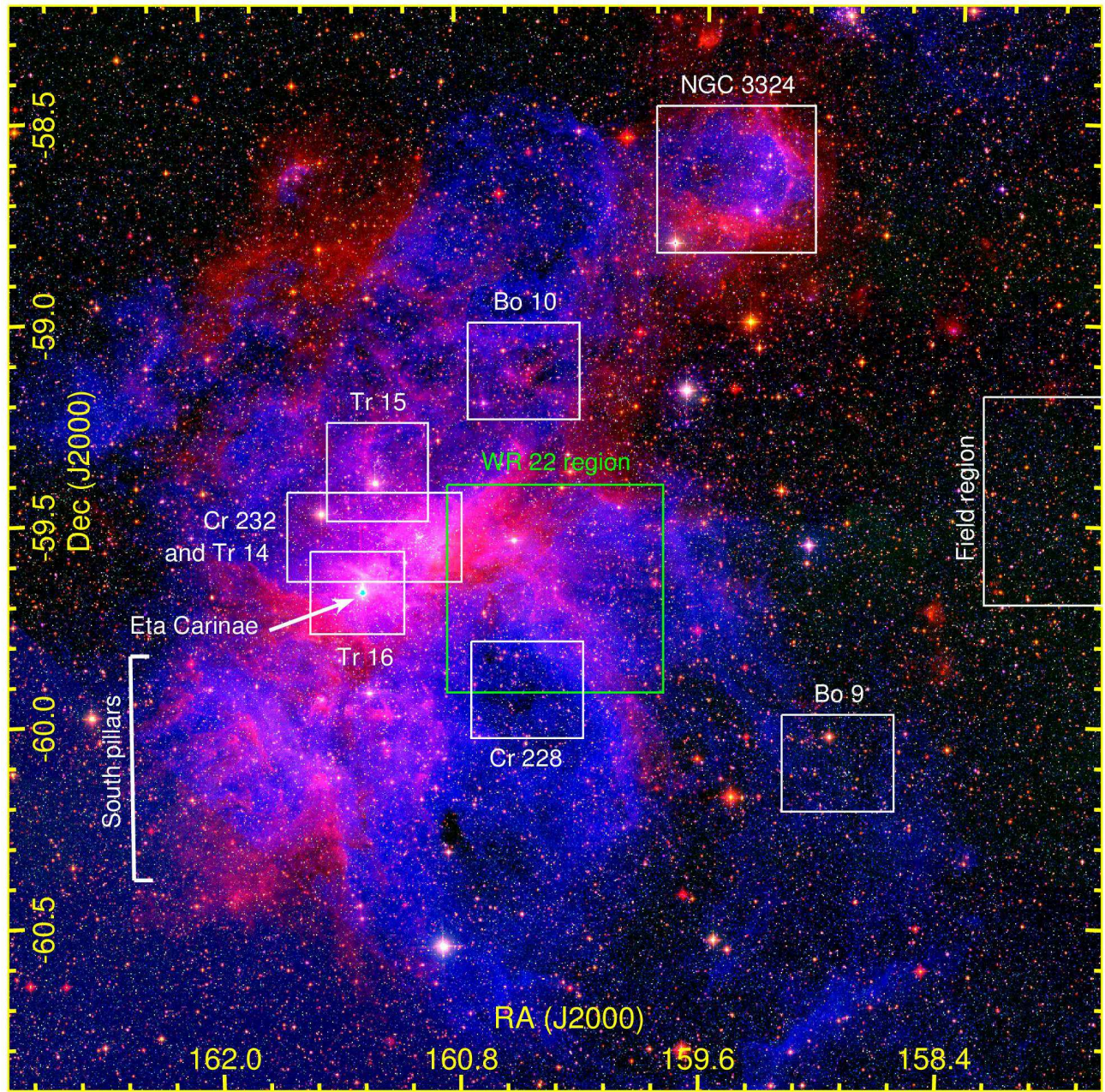}
\caption{\label{totalregion} Color composite image of the large ($2.7^{\circ} \times 2.7^{\circ}$)
area containing the Carina Nebula and centered at $\alpha$(J2000)~=~10$^{\rm h}$~41$^{\rm m}$~17\fs5
and $\delta (J2000) = -59\degr ~40\arcmin ~36\farcs9$.
This RGB image was made using the {\rm WISE} 4.6 $\mu$m (red), 2MASS $K_s$ band (green),
and {\rm DSS} $R$ band (blue) images. Approximate locations of different star clusters
(Tr~14, 15, 16; Bo~9, 10; Cr~228, 232, and NGC~3324) are denoted by white boxes.
$\eta$ Carinae is marked by an arrow and in the lower left part of the image,  
south pillars \citep{2000ApJ...532L.145S} are shown. The region covered in the present 
study is shown by the green box. Part of the selected field region can be seen in the 
extreme western part of the image. North is up and east is to the left.}
\end{figure*}

Initially, on the basis of infrared (IR) and molecular studies of the central Carina region, 
several authors \citep[see][]{1979ApJ...227..114H, 1988ApJ...330..928G, 1981AA...102..257D, 
1995RMxAC...2..105C} have reported that the Carina nebula is an evolved H{\sc ii} region and that 
there is a paucity of active star formation. However, following the detection of several 
embedded IR sources, \citet{2000ApJ...532L.145S} showed that star formation is still going on 
in this region. Later, \citet{2001MNRAS.327...46B} also identified two compact H{\sc ii} 
regions possibly linked with very young O-type stars. \citet{2002MNRAS.331...85R} traced the 
photodissociation regions (PDRs) that are expected to be present in the massive star-forming 
regions. They conclude that the star formation within the Carina region has certainly not been 
completely halted despite prevailing unfavorable conditions imposed by the very hot massive stars 
\citep[see for more details][]{Claeskens2011}. Detection of proplyds-like objects in these regions 
\citep[see][]{1998RMxAC...7R.217D, 2003ApJ...587L.105S} proves the ongoing active low- and 
intermediate-mass star formation. Very recently, an isolated neutron star candidate discovered in 
the neighborhood of $\eta$ Carinae suggests there are at least two episodes of massive 
star formation \citep{2007PASJ...59S.151H, 2009AA...498..233P}.

Figure~\ref{totalregion} shows a three-color composite image, using the {\rm WISE} 4.6 $\mu$m, 
2MASS $K_s$ band and {\rm DSS} $R$ band images, of the large region of the 
Carina nebula. The prominent V-shaped lane is associated with the nebular complex and consists 
of dust and molecular gas \citep{1974AA....31...11D}. Trumpler (Tr) 16 is located near the 
central portion of this lane and thought to be $\sim$3 Myr old. This cluster also hosts one of 
the most massive stars in our galaxy, $\eta$ Carinae (indicated by an arrow in the image), which 
has an estimated initial mass $\gtrsim$ 150 M$_{\odot}$ \citep{2001ApJ...553..837H}. Tr~14 is 
younger with an age of $<$ 2\,Myr \citep{2004AA...418..525C, 2008hsf2.book..138S}. 
In between Tr~14 and Tr~16, there is another cluster named Collinder (Cr) 232. The cluster Cr~228 
near Tr~16 is very young and probably located in front of the Carina nebula complex 
\citep{2001AA...379..136C}. Finally NGC~3324 (upper part in the image) is believed to be located 
inside the Carina spiral arm and embedded in a filamentary elliptical shaped nebulosity 
\citep[see][]{2001AA...371..107C}. 

Since the Carina nebula is a typical star-forming region, feedback from the young and massive 
stars has cleared out the nebulosity in the central region and a large number of elongated 
structures, so-called Pillars (\citealt{2000ApJ...532L.145S}, shown in the lower left part of 
the image) have formed in the outer regions. We can see many of them in the southern part of 
the image. We also observe large bubbles in the northern region, probably caused by the 
gusts of hot gas leaking from the powerful stars at the center of the nebula
\citep{2000ApJ...532L.145S}. The central clusters Tr~14 and Tr~16 tend to be devoid of 
star formation \citep{2008hsf2.book..138S}, but there are  active sites of ongoing star 
formation in the outer regions of the nebula. In the present study, our aim is to 
understand the star formation in one of the peripheral regions of the Carina nebula, 
influenced by the presence of hot massive stars. 

Because of its relatively low obscuration and proximity and its rich stellar content, this 
nebula is one of the most extensively explored nearby objects \citep{2008hsf2.book..138S}. 
Several wide-field surveys of the Carina Nebula complex (CNC) have recently been carried 
out at different wavelengths. The combination of a large {\it Chandra} X-ray survey 
\citep[see][]{2011ApJS..194....1T} with a deep near-infrared (NIR) survey 
\citep{2011AA...530A..34P, 2011AA...525A..92P}, Spitzer mid-infrared (MIR) observations 
\citep{2010MNRAS.406..952S, 2011ApJS..194...14P}, and Herschel far-infrared (FIR) observations 
\citep{2013AA...549A..67G, 2013AA...554A...6R} provides comprehensive information about the 
young stellar populations. In this paper, we discuss our new optical photometry, along 
with some low resolution spectroscopy, archival NIR (2MASS), and X-ray ({\it Chandra}, 
{\it XMM-Newton}) data of a field located west of $\eta$ Carinae (hereafter CrW) 
and centered on the WN7ha + O binary system WR~22 \citep[HD 92740;][]{1979ApJ...228..206C, 
1979IAUS...83..291N, 1981SSRv...28..227V, 1991IBVS.3571....1G, 1991AA...249..443H, 
1995AA...293..403C, 1996AA...306..771R, 2009AA...508..805G} 
positioned just outside the V-shaped dark lane.

\section{Observations and data analysis}

\subsection{Optical photometry}
A set of  $UBVRI$ and $H\alpha$ observations of CrW
($\alpha$(J2000)~=~10$^{\rm h}$~41$^{\rm m}$~17\fs5
and $\delta (J2000) = -59\degr ~40\arcmin ~36\farcs9$)
were obtained with the Wide Field Imager (WFI) instrument at the ESO/MPG 2.2 m 
telescope at La Silla in March 2004 (service mode, 72.D-0093 PI: E. Gosset). The WFI 
instrument has a field of view of about $34' \times 33'$, covered by a mosaic of eight CCD 
chips with a pixel size of 0.238 arcsec. The observations typically consisted of 
three dithered frames with a short exposure time (about 50s in $U$, 10s in $BVRI$, and 100s 
in $H\alpha$) and three dithered frames with about 18 times longer exposures to allow measurements 
of both bright and faint objects. Additional frames of a field located closer to 
the main Carina region were also acquired in order to connect our photometric system
to those of previous works.

The data were bias-subtracted, flat-fielded and corrected for cosmic-rays using the standard tasks 
available in {\it IRAF.\,}\footnote{{\it IRAF} (Image Reduction and Analysis Facility) is distributed
by the National Optical Astronomy Observatories, which is operated by the Association of Universities for 
Research in Astronomy, Inc. under co-operative agreement with the National Science Foundation.} The 
photometry in the natural system was obtained with the {\it DAOPHOT}\footnote{{\it DAOPHOT} stands for 
Dominion Astrophysical Observatory Photometry.} \citep{1987PASP...99..191S, 1992ASPC...25..297S} software.
We also performed aperture photometry of Stetson's and Landolt's standard fields and of the additional 
frames. All of them, along with ESO recommendations, were used to determine the color transformation 
coefficients. The zero points were fixed via comparison with data published by \citet{1993AJ....105..980M}, 
\citet{1996AAS..116...75V}, \citet{2001ApJ...549..578D} and mainly with the unpublished catalog of 
\citet{2003MNRAS.339...44T}.

The following equations were adopted, together with appropriate zero points:

\[V_{std}           = V_{wfi}             - 0.107 * (B-V)_{wfi} , \]
\[(B-V)_{std}       =                      1.440 *  (B-V)_{wfi} , \]
\[(U-V)_{std}       = 1.08 * (U-V)_{wfi}   + 0.02 * (B-V)_{wfi} , \]
\[(V-R)_{std}       = 0.98 * (V-R)_{wfi}   - 0.09 * (B-V)_{wfi} , \]
\[(V-I)_{std}       = 0.94 * (V-I)_{wfi}   - 0.08 * (B-V)_{wfi} . \]

The color transformation coefficients and the zero points obtained above were then used further 
to calibrate the aperture photometry of 50 well-isolated bright sources in the CrW region. 
The astrometry was established by matching the instrumental coordinates with the 2MASS point 
source catalog. The rms of the astrometric calibration is 0.15$\arcsec$ in RA and 0.19$\arcsec$ 
in Dec. To avoid source confusion due to crowding, PSF (point spread function) photometry 
was collected for all the sources in the CrW region. PSF photometric magnitudes were generated by 
the {\it ALLSTAR} task inside the {\it DAOPHOT} package. The calibrated aperture magnitudes of 
the same 50 stars were then used to calibrate the magnitudes of all the stars in the CrW region 
obtained from the PSF photometry.

These final PSF calibrated magnitudes were used in further analysis. The typical {\it DAOPHOT} 
errors are found to increase with the magnitude and become large ($\ge$ 0.1 mag) for stars fainter 
than $V$ $\ge$ 22 mag. The measurements beyond this magnitude were not considered in our analysis.
In addition, for the present study, we used only the $32\arcmin \times 31\arcmin$ 
inner area of the mosaic.

\subsection{Completeness of the data}\label{cod}
\begin{figure}
\centering\includegraphics[height=6.5cm,width=8.5cm,angle=0]{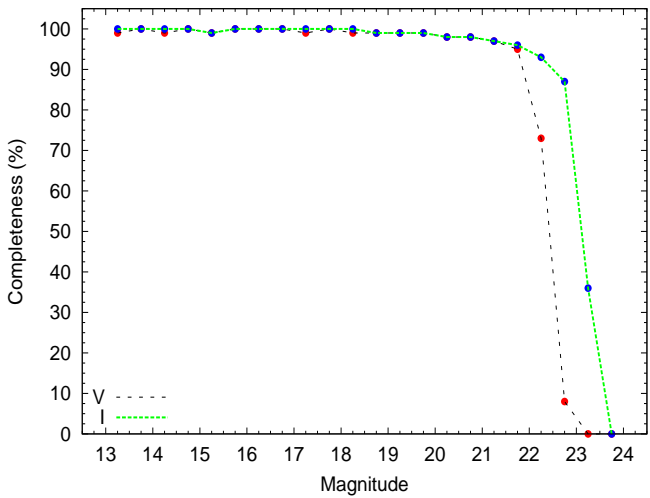}
\caption{\label{cft} Completeness levels for the $V$ and $I$ bands as a function of magnitude
derived from an artificial star experiment ({\it ADDSTAR}, see Sect.~\ref{cod}).}
\end{figure}

There could be various reasons (e.g., crowding of the stars) that the completeness 
of the data sample may be affected. Establishing the completeness is very important 
to study the luminosity function (LF)/mass function (MF). The {\it IRAF} 
routine {\it ADDSTAR} of {\it DAOPHOT {\sc II}} was used to determine the completeness 
factor (CF). Briefly, in this method, artificial stars of known magnitudes and positions 
from the original frames are randomly added, and then artificially generated frames 
are reduced again by the same procedure as used in the original reduction. The ratio 
of the number of stars recovered to those added in each magnitude gives the CF as a 
function of magnitude. In Fig.~\ref{cft}, we show the CF as a function of the $V$ magnitude. As 
expected, the CF decreases as magnitude increases. Our photometry is more than 90\% complete 
up to $V$ = 21.5 and $I$ = 22 magnitude. For the distance of 2.9 kpc (cf. Sect.~\ref{distance}), 
this will limit our study to pre-main-sequence (PMS) stars more massive than 0.5 M$_\odot$.

\begin{figure}
\centering\includegraphics[height=8cm,width=8cm,angle=0]{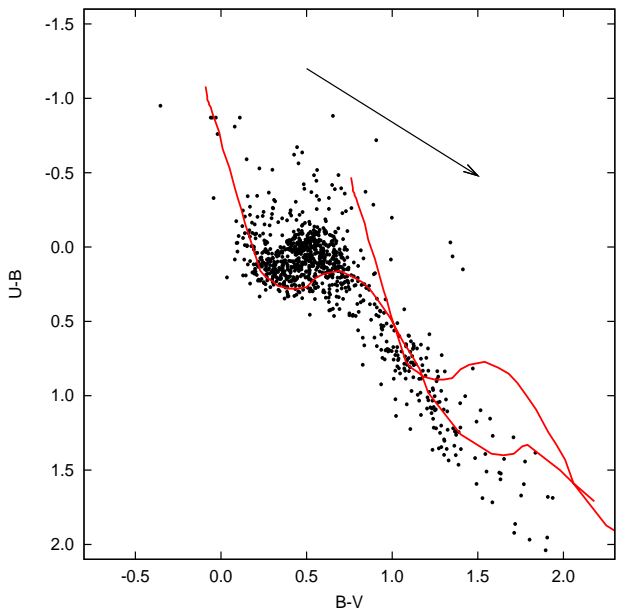}
\caption{\label{ccopt} $(U-B)/(B-V)$ two color diagram for all the stars lying in the
region with $V<16$ mag. The two continuous curves represent the ZAMS by \citet{Schmidt-Kaler1982}
shifted for the minimum ($E(B-V)$ = 0.25, left) and maximum ($E(B-V)$ = 1.1, right)
reddening values. The reddening vector with a slope of 0.72 and size of $A_v$ = 3 mag is also 
shown.}
\end{figure}

\subsection {Spectroscopy}\label{spect} 
For a set of 15 X-ray sources\footnote{Throughout this paper, we used the numbering convention
of X-ray sources as introduced in \citet{Claeskens2011}.} identified using {\it XMM-Newton} observations 
in the CrW field \citep[see][]{Claeskens2011}, we obtained their optical spectra between 4 and 6 
March 2003 using the EMMI instrument mounted on the ESO 3.5 m New Technology Telescope (NTT) at La 
Silla (PI: E. Gosset). This instrument was used in the Red Imaging and Low Dispersion Spectroscopy 
(RILD) mode with grism $\#5$ (wavelength range 4000 - 8700 \AA). One spectrum was obtained with the 
VLT + FORS1 \citep[see][]{Claeskens2011}. The data were reduced in the standard way using the 
{\tt long} context of the {\rm ESO-{\it MIDAS}} (European Southern Observatory Munich Image Data 
Analysis System) package\footnote{{\rm ESO-{\it MIDAS}} has been developed by the European Southern 
Observatory.}. Since the observing conditions were favorable during our run, target spectra were 
calibrated using the flux spectrum of the standard star LTT~2415 \citep{1992PASP..104..533H}. 

\subsection{Archival data: 2MASS}

We used the 2MASS Point Source Catalog (PSC) \citep{2003yCat.2246....0C} for NIR ($JHK_s$) 
photometry of point sources in the CrW region. This catalog is said to be 99$\%$ complete up 
to the limiting magnitudes of 15.8, 15.1 and 14.3 in the $J$ (1.24$\mu$m), $H$ (1.66$\mu$m), and 
$K_s$ (2.16$\mu$m) bands, respectively\footnote{http://tdc-www.harvard.edu/catalogs/tmpsc.html}.
We selected only those sources that have NIR photometric accuracy $<$ 0.2 mag and detection 
in at least the $K_s$ and $H$ bands. Since the seeing ($\sim$FWHM of the stars intensity profile)
for the WFI observations was around 1 arcsec, the optical counterparts of the 2MASS sources were 
searched using a matching radius of 1 arcsec.

\section{Basic parameters}

\subsection{Reddening}\label{reddening}

The $(U - B)/(B - V)$ two-color diagram (TCD) was used to estimate the extinction 
toward the CrW region. In Fig.~\ref{ccopt}, we show the TCD with the zero-age-main-sequence 
(ZAMS) from \citet{Schmidt-Kaler1982} shifted along the reddening vector with a slope of 
$E(U - B)/E(B - V)$ = 0.72 to match the observations. This shift will give the extinction 
directly toward the observed CrW region. The distribution of stars shows a wide 
spread in the diagram along the reddening line indicating the clumpy nature of the molecular 
cloud associated with this star-forming region. If we look at the MIR image of CrW (for detail see 
Sect.~\ref{diss} and Fig.~\ref{spa}), we see the dark dust lane along with several enhancements 
of nebular materials at many places that are likely to be responsible for this spread in 
reddening.
Figure~\ref{ccopt} yields a minimum reddening value $E(B-V)$ of 0.25 with a wide spread leading
to values up to 1.1 mag. Recent works (see Table~\ref{earlyob}) also indicate a spread in 
the value of $E(B - V)$ ($\sim$0.3 $-$ 0.8 mag) toward the $\eta$ Carinae region.
\citet{2008hsf2.book..138S} suggest that a detailed optical study of the Carina 
nebula can easily be done since our sight line toward this nebula suffers little extinction 
and reddening compared to most of the massive star-forming regions. This seems true for the 
line of sight up to the first stars belonging to the complex, but it could perhaps not remain 
applicable to objects farther away and embedded inside the molecular cloud.

\subsection{Reddening law}\label{rdl}

\begin{figure*}
\centering\includegraphics[height=14cm,width=15.5cm,angle=0]{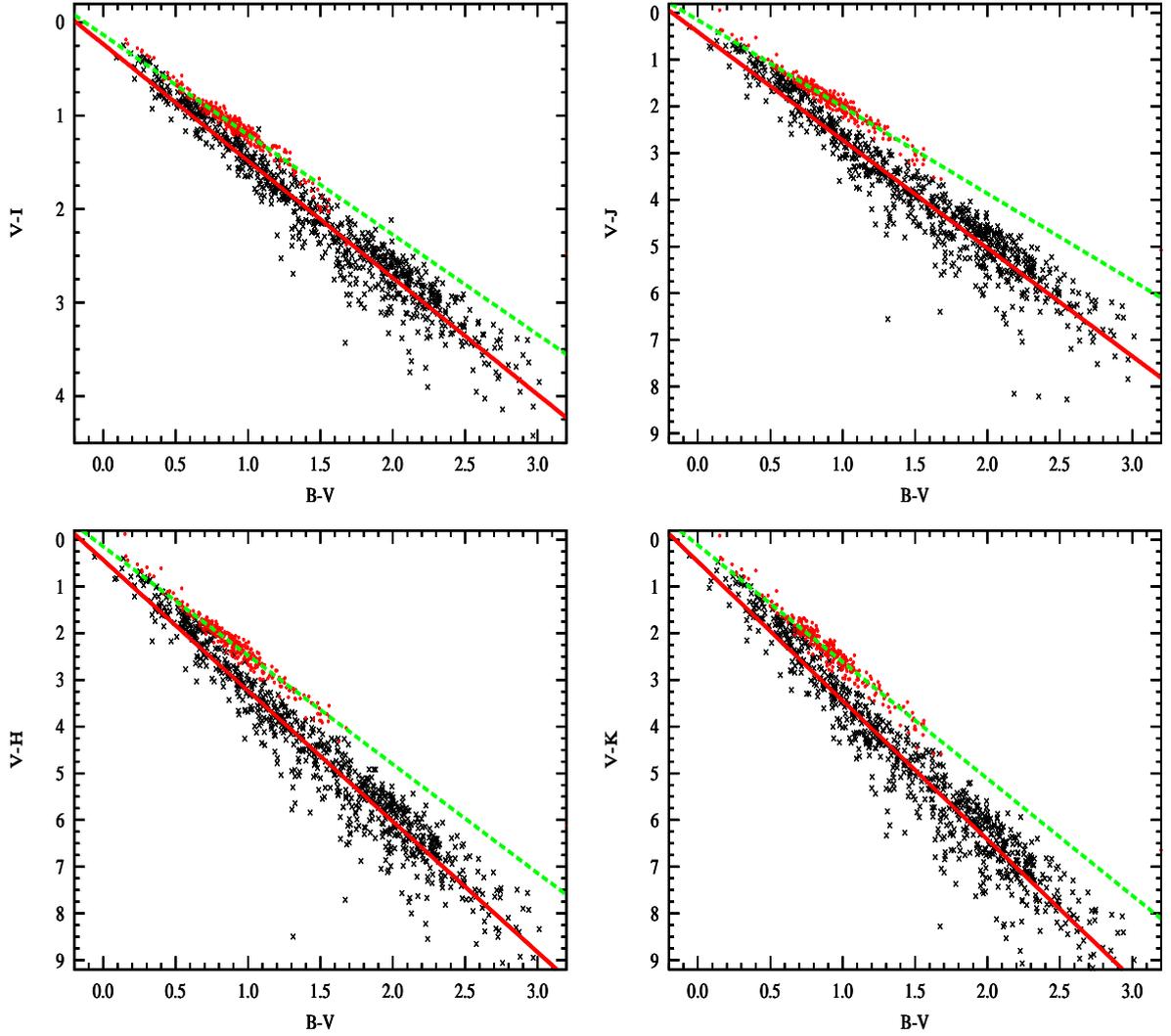}
\caption{\label{2color} $(V-I),(V-J), (V-H)$, and $(V-K)$ versus $(B-V)$ TCDs
for the stars in the CrW region ($r <10\arcmin$ from WR~22). The cross and dot symbols
represent the stars with abnormal and normal reddening, respectively.
Straight and dotted lines show least-squares fits to the data.}
\end{figure*}

To study the nature of the gas and dust in young star-forming regions, it is very important to 
know the properties of the interstellar extinction and the ratio of total-to-selective extinction, 
i.e., $R_V = A_{V}/E(B - V)$. The normal reddening law for the solar neighborhood has been estimated
to be $R_V$ = 3.1 $\pm$ 0.2 \citep[cf.][]{1989AJ.....98..611G, 2003dge..conf.....W,
2011JKAS...44...39L} but in the case of the $\eta$ Carinae region, there are several studies
that claim that $R_V$ is anomalously high \citep[see][]{1973AAS...12..331F, 1976ApJ...208..923H, 
1978AJ.....83.1199F, 1980AA....89..209T, 1987MNRAS.227..943S, 2002MNRAS.331....7S, 
1988MNRAS.232..661T, 1996AAS..116...75V}. Recently, using 141 early type members in this 
region, \citet{2012AJ....143...41H} derived an abnormal total-to-selective extinction ratio 
$R_V$ = 4.4 $\pm$ 0.04. 

\begin{table*}
\centering
\scriptsize
\caption{Extinction, distance, and reddening values for the Carina region collected from the literature.}
\begin{tabular}{lllllll} \hline \hline
Region/cluster& $E(B - V)$     & $R_{V}$         & $M_{0}-M{_V}$  & d(kpc)      & References                  & Method/techniques      \\
\hline
WR~22 (CrW)  & 0.36            & 3.1           & 12.15          & 2.7           & \citet{2009AA...508..805G} & --                     \\
\hline
Trumpler 14  &       --        &       --      & 11.1           &       --      & \citet{1971AAS....4..241B} & --                     \\
             &       0.5       &       --      & 11.5           & 2.0           & \citet{1971AA....14..120T} & --                     \\
             &       --        &       --      & 12.7 $\pm$ 0.2 & 3.5           & \citet{1973ApJ...179..517W} & Spectroscopic parallax \\
             &       --        & 3.2           & 12.7 $\pm$ 0.1 &       --      & \citet{1978ApJS...38..309H} & --                     \\
             &       --        &       --      & --             & 2.8           & \citet{1980AA....89..209T} & --                     \\
             &       --        &       --      & 12.3 $\pm$ 0.1 &       --      & \citet{1982AJ.....87.1300W} & --                     \\
             & 0.55 $\pm$ 0.08 &       --      & 12.2 $\pm$ 0.2 &       --      & \citet{1983ApSS..96..293F} & --                     \\
             &       --        & 3.2           & 12.7 $\pm$ 0.1 & 3.5           & \citet{1988PASP..100.1431M} & Spectroscopic parallax \\
             &       --        &       --      & 12.3 $\pm$ 0.1 & 2.8           & \citet{1988PASP..100.1431M} & Spectroscopic parallax \\
             & 0.82 $\pm$ 0.12 &       --      & 11.9 $\pm$ 0.2 & 2.4 $\pm$ 0.3 & \citet{1988MNRAS.232..661T} & Main-sequence fitting  \\
             &       --        & 3.2           & 12.8 $\pm$ 0.2 &       --      & \citet{1993AJ....105..980M} & Spectroscopic parallax \\
             & 0.57 $\pm$ 0.13 & 4.7 $\pm$ 0.7 & 12.5 $\pm$ 0.2 & 3.1 $\pm$ 0.3 & \citet{1996AAS..116...75V} & Main-sequence fitting  \\
             & 0.58            & 3.2           & 12.8 $\pm$ 0.1 &       --      & \citet{2001ApJ...549..578D} & Spectroscopic parallax \\
             &       --        &       --      & 12.2 $\pm$ 0.7 &       --      & \citet{2003MNRAS.339...44T} & Spectroscopic parallax \\
             & 0.57 $\pm$ 0.12 & 4.2 $\pm$ 0.2 & 12.0 $\pm$ 0.2 & 2.5 $\pm$ 0.3 & \citet{2004AA...418..525C} & Main-sequence fitting  \\
             & 0.36 $\pm$ 0.04 & 4.4 $\pm$ 0.2 & 12.3 $\pm$ 0.2 & 2.9 $\pm$ 0.3 & \citet{2012AJ....143...41H} & Proper motion          \\
\hline
Trumpler 15  &       --        &       --      & 11.1           &       --      & \citet{1971AAS....4..241B} & --                     \\ 
             &       0.4       &       --      & 11.5           & 1.6           & \citet{1971AA....14..120T} & --                     \\
             &       --        &       --      & 12.9           & 3.7           & \citet{1973ApJ...179..517W} & Spectroscopic parallax \\
             &       --        &       --      & 12.9           &       --      & \citet{1978ApJS...38..309H} & --                     \\
             &       --        &       --      & --             & 2.5           & \citet{1980AA....89..209T} & --                     \\
             &       --        & 3.2           & 11.8 $\pm$ 0.1 & 2.3           & \citet{1988PASP..100.1431M} & Spectroscopic parallax \\
             &       --        &       --      & 12.1 $\pm$ 0.3 & 2.6           & \citet{1988PASP..100.1431M} & Spectroscopic parallax \\
             & 0.49$\pm$ 0.09  &       --      & 12.1 $\pm$ 0.2 & 2.6 $\pm$ 0.3 & \citet{1988MNRAS.232..661T} & Main-sequence fitting  \\
             & 0.49$\pm$ 0.09  &       --      & 12.1 $\pm$ 0.2 & 2.9           & \citet{2003MNRAS.339...44T} & Spectroscopic parallax \\
             &       --        &       --      & 12.3 $\pm$ 0.2 &       --      & \citet{2004AA...418..525C} & Main-sequence fitting  \\
\hline
Trumpler 16  & 0.44            &       --      & 11.9           & 2.5           & \citet{1971AA....14..120T} & --                     \\
             & 0.4             &       --      & 12.1           & 2.7           & \citet{1973AAS...12..331F} & --                     \\
             &       --        & 3.0           & 12.1 $\pm$ 0.2 & 2.6           & \citet{1973ApJ...179..517W} & Spectroscopic parallax \\
             &       --        &       --      & 12.2 $\pm$ 0.1 &       --      & \citet{1978ApJS...38..309H} & --                     \\
             &       --        &       --      & 12.0           & 2.8           & \citet{1980AA....89..209T} & --                     \\
             &       --        & 3.1           & 11.8 $\pm$ 0.1 & 2.3           & \citet{1982PASP...94..807L} & Spectroscopic parallax \\
             &       --        &       --      & 12.3 $\pm$ 0.1 &       --      & \citet{1982AJ.....87.1300W} & --                     \\
             & 0.68 $\pm$ 0.15 &       --      & 12.0 $\pm$ 0.2 & 2.5 $\pm$ 0.2 & \citet{1988MNRAS.232..661T} & Main-sequence fitting  \\
             &       --        &       --      & 12.5 $\pm$ 0.1 &       --      & \citet{1993AJ....105..980M} & Spectroscopic parallax \\
             & 0.58            & 3.2           & 12.8 $\pm$ 0.1 &       --      & \citet{2001ApJ...549..578D} & Spectroscopic parallax \\
             &       --        &       --      & 12.0 $\pm$ 0.6 & 2.5           & \citet{2003MNRAS.339...44T} & Spectroscopic parallax \\
             & 0.61 $\pm$ 0.15 & 3.5 $\pm$ 0.3 & 13.0 $\pm$ 0.3 & 3.9 $\pm$ 0.5 & \citet{2004AA...418..525C} & Main-sequence fitting  \\
             & 0.36 $\pm$ 0.04 & 4.4 $\pm$ 0.2 & 12.3 $\pm$ 0.2 & 2.9 $\pm$ 0.3 & \citet{2012AJ....143...41H} & Proper motion          \\
\hline
Collinder 228&       --        &       --      & 12.0 $\pm$ 0.2 & 2.5           & \citet{1973AAS...12..331F} & --                     \\
             &       --        &       --      & 12.2           &       --      & \citet{1973ApJ...179..517W} & Spectroscopic parallax \\
             &       --        &       --      & 12.0 $\pm$ 0.3 &       --      & \citet{1978ApJS...38..309H} & --                     \\
             &       --        & 3.2           &       --       & 2.5           & \citet{1980AA....89..209T} & --                     \\
             &       --        &       --      & 12.06          & 2.6           & \citet{1981PASP...93..714L} & Spectroscopy           \\
             & 0.64 $\pm$ 0.26 &       --      & 11.6 $\pm$ 0.4 & 2.1 $\pm$ 0.4 & \citet{1988MNRAS.232..661T} & Main-sequence fitting  \\
             &       --        &       --      &       --       & 1.9 $\pm$ 0.2 & \citet{2001AA...379..136C} & --                     \\
\hline
Collinder 232& 0.68 $\pm$ 0.21 &       --      & 12.0 $\pm$ 0.2 & 2.5 $\pm$ 0.2 & \citet{1988MNRAS.232..661T} & Main-sequence fitting  \\
             & 0.48 $\pm$ 0.12 & 3.7$\pm$ 0.03 & 11.8 $\pm$ 0.2 & 2.3 $\pm$ 0.3 & \citet{2004AA...418..525C} & Main-sequence fitting  \\
\hline
Bochum 9     & 0.63 $\pm$ 0.08 &       --      &       --       & 4.7           & \citet{2001MNRAS.325.1591P} & --                     \\
\hline
Bochum 10    &       --        &       --      & 12.8           &       --      & \citet{1981PASP...93..202F} & --                     \\
             &       --        &       --      & 12.2           &       --      & \citet{1987MNRAS.228..545F} & --                     \\
             & 0.48 $\pm$ 0.05 &       --      & 12.2           & 2.7           & \citet{2001MNRAS.325.1591P} & --                     \\
\hline
NGC 3324     &       --        &       --      & 12.5 $\pm$ 0.2 &      --       & \citet{1977AAS...27..145C} & --                     \\
             &       --        &       --      & 12.4 $\pm$ 0.03& 3.0 $\pm$ 0.1 & \citet{2001AA...379..136C} & --                     \\
\hline
Trumpler 14, &       --        & 3.2 $\pm$ 0.3 & 12.2           & 2.7 $\pm$ 0.2 & \citet{1980AJ.....85.1193T} & --                     \\
15, 16 and Cr 228&             &               &                &               &                             &                        \\
\hline
\end{tabular} \\
\label{earlyob}
\end{table*}

We used the TCDs as described by \citet{2003AA...397..191P} to study the nature of the 
extinction law in the CrW region. The TCDs of the form of $(V - \lambda)$ versus $(B - V)$, 
where $\lambda$ indicates one of the wavelengths of the broad-band filters ($R, I, J, H, K, L$), 
provide an effective method for distinguishing the influence of the normal extinction produced by 
the diffuse interstellar medium from that of the abnormal extinction arising within regions 
having a peculiar distribution of dust sizes \citep[cf.][]{1990AA...227..213C, 2000PASJ...52..847P}.

We clearly see in Fig.~\ref{2color} that there are two types of distribution having 
different slopes. We selected all the stars belonging to these two populations and 
plotted their $(V-I), (V-J), (V-H)$ and $(V-K)$ vs. $(B-V)$ TCDs in Fig.~\ref{2color}.
The respective slopes relating these colors were found, for the red-dot stars,
to be $1.07\pm0.02, 1.86\pm0.02, 2.33\pm0.03$, and $2.50\pm0.03$, which are approximately 
equivalent to the normal galactic values, i.e., 1.10, 1.96, 2.42, and 2.60, respectively.
The objects with black crosses display steeper slopes, i.e., $1.28\pm0.01, 2.34\pm0.03, 
2.84\pm0.03$ and $3.03\pm0.03$ for $(V-I), (V-J), (V-H)$ and $(V-K)$ vs. $(B-V)$, 
respectively.
If we plot the spatial distribution of the red dots and black crosses, we clearly see that 
all the red dots are uniformly distributed, whereas all the black crosses are distributed 
away from the obscured region of the molecular cloud. It means that the black crosses are 
most probably background stars, and their light is seen through the molecular cloud 
\citep[see][and references therein]{2011ApJS..194...10P, 2013AA...554A...6R}. 
Therefore, the ratios [$E(V - \lambda)$]/[$E(B - V)$] ($\lambda$ $\geq$ $\lambda_I$) for 
the stars in the background yield a high value for $R_V$ ($\sim$3.7 $\pm$ 0.1), indicating an 
abnormal grain size in the observed region. Many investigators (see column 3 of 
Table~\ref{earlyob}) have also found evidence of larger dust grains 
in the Carina region. \citet{1993AJ....105..258M} have found that the value of $\lambda_{max}$ 
(the wavelength at which maximum polarization occurred, which is also an indicator of the 
mean dust grain size distribution) is higher than the canonical value for the general 
diffuse ISM.

Several studies have already pointed toward an anomalous reddening law with a high $R_V$ value in 
the vicinity of star-forming regions \citep[see, e.g.,][]{2003AA...397..191P}.
However, for the Galactic diffuse interstellar medium, a normal value of $R_V$ = 3.1 is well 
accepted. The higher-than-normal value of $R_V$ has usually been attributed to the presence 
of larger dust grains. There is evidence that, within the dark clouds, accretion of ice 
mantles on grains and coagulation due to colliding grains change the size distribution
towards larger particles. On the other hand, in star-forming regions, radiation from 
massive stars may evaporate ice mantles resulting in small particles. 
Here, it is interesting to mention that \citet{2003AA...412..199O} suggest that 
efficient dust destruction is undergoing in the ionized region on the basis of the
[Si II] 35 to [N II] 122 $\mu$m ratio. \citet{1983AA...117..289C} and 
\citet{1990AA...227..213C} have shown that both larger and smaller grains may 
increase the ratio of total-to-selective extinction.

\begin{figure*}
\centering\includegraphics[height=8cm,width=9cm,angle=0]{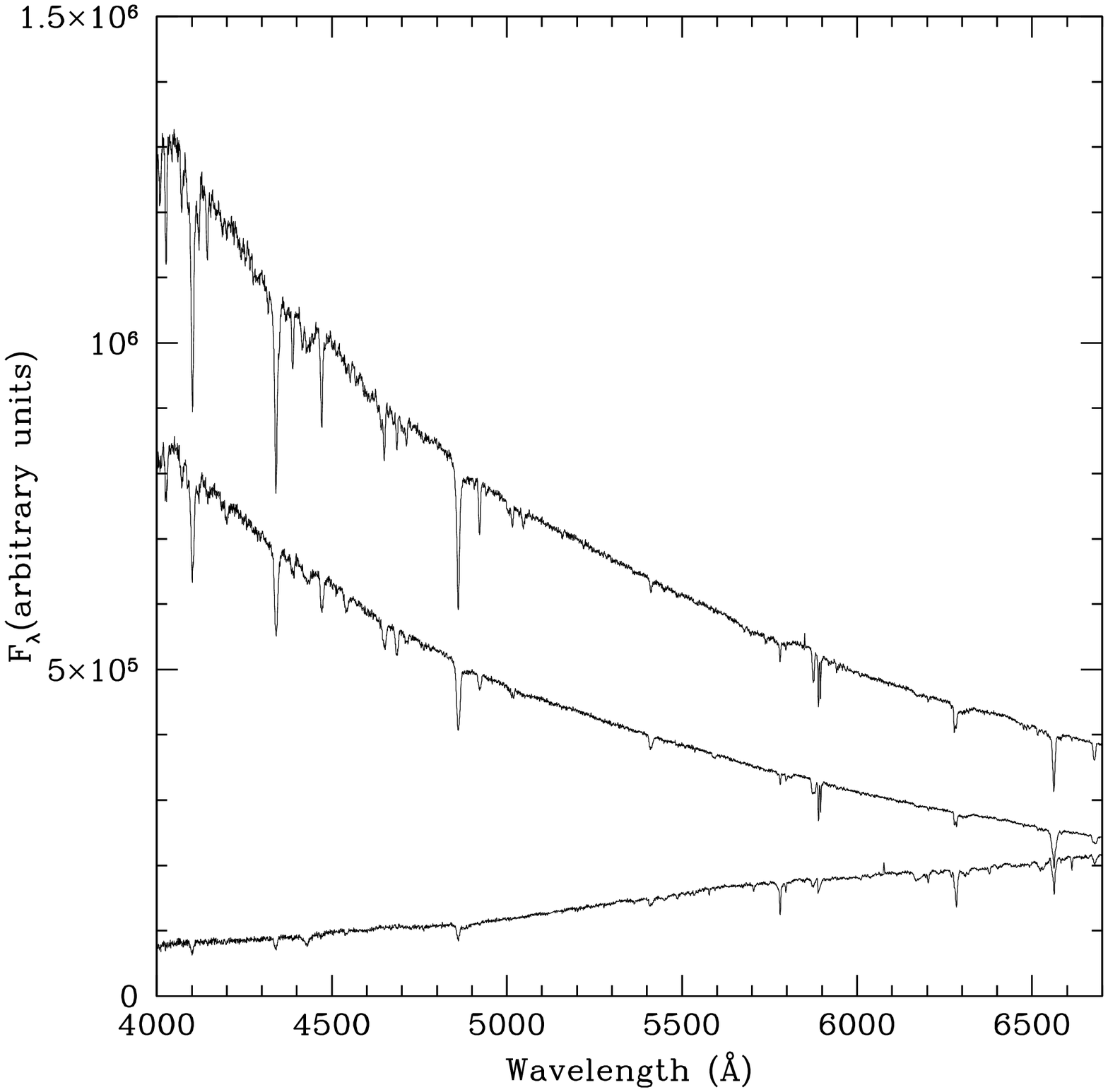}
\centering\includegraphics[height=8cm,width=9cm,angle=0]{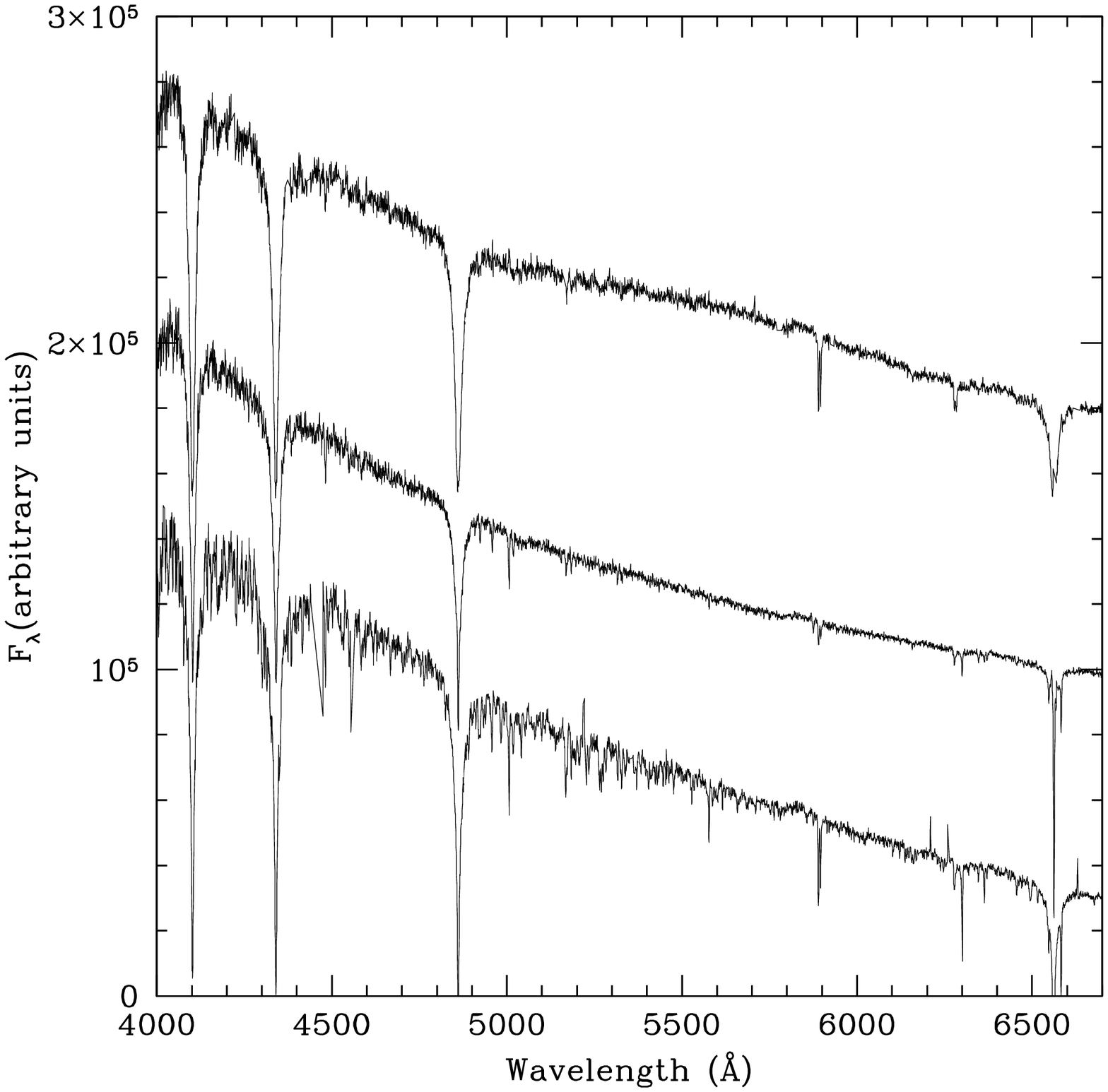}
\centering\includegraphics[height=8cm,width=9cm,angle=0]{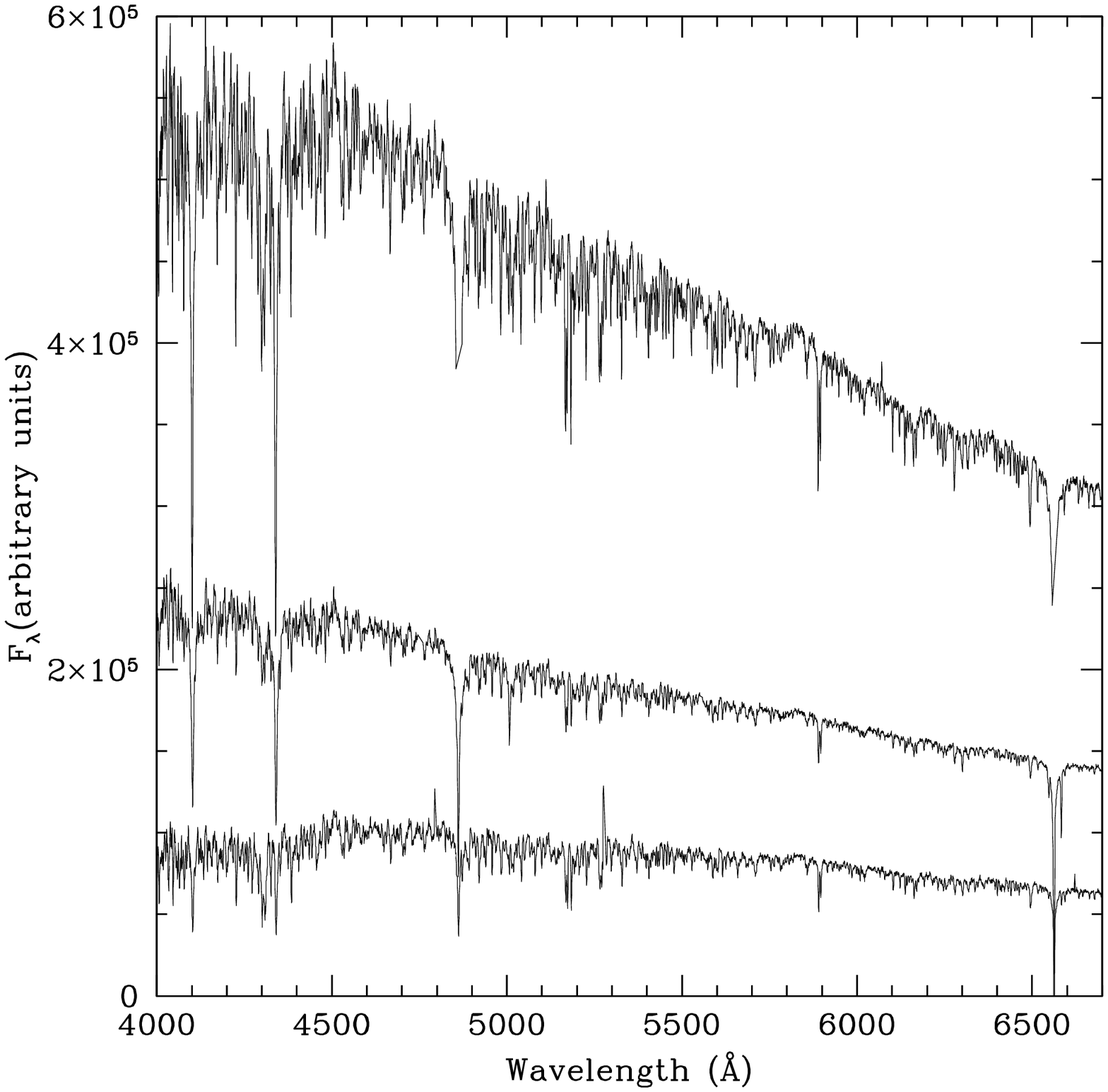}
\centering\includegraphics[height=8cm,width=9cm,angle=0]{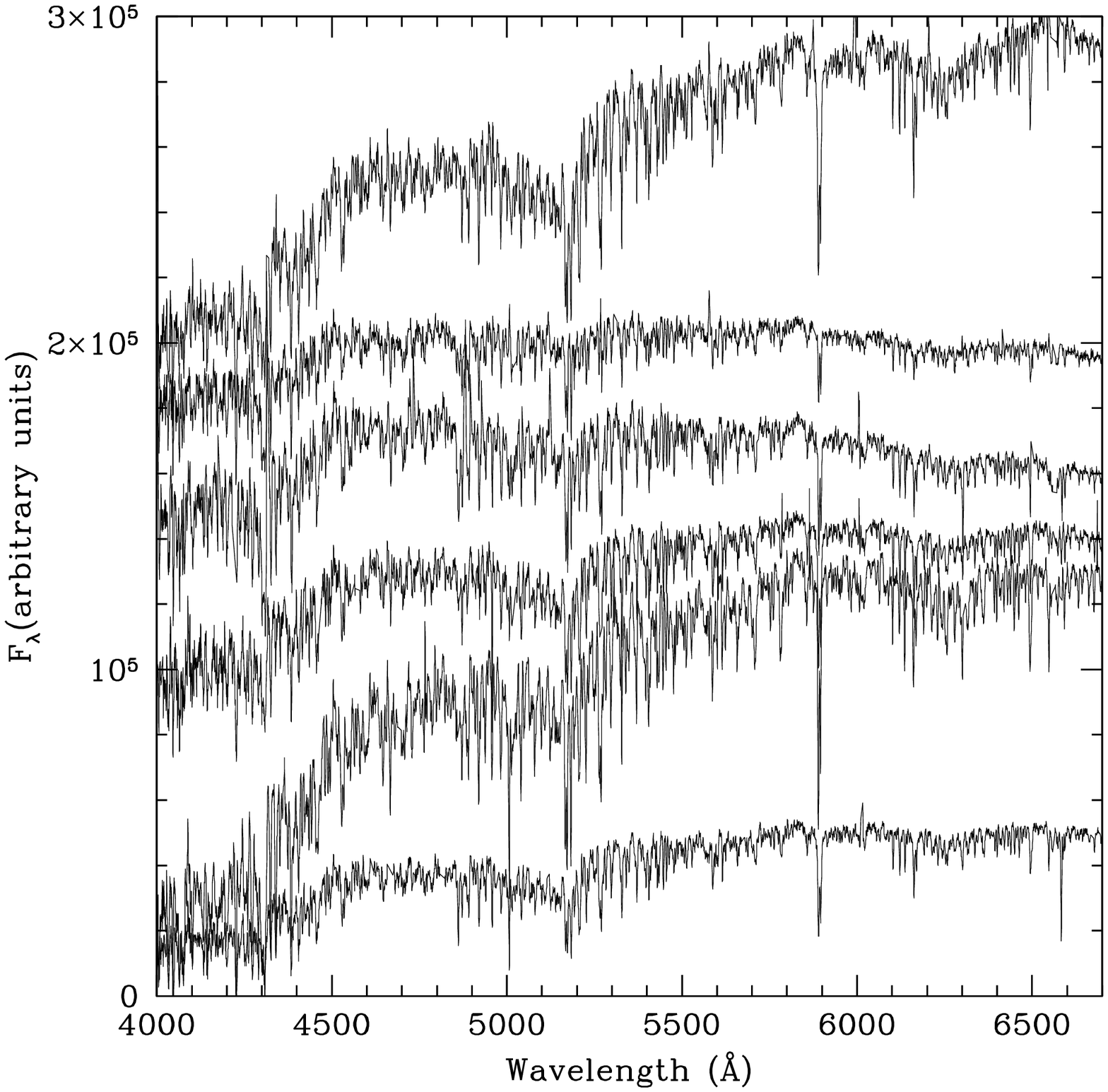}
\caption{Flux-calibrated spectra of the O-A-F-G type stars in our spectroscopic sample
of the CrW region. The spectra have been randomly shifted vertically for clarity. The
spectral types become progressively later from left to right and from top to bottom.}
\label{fig-spec}
\end{figure*}

\subsection{Distance}\label{distance}

The Carina nebula is a very large (angular size $>$ $2^{\circ} \times 1.5^{\circ}$)
active star-forming region containing a number of young star clusters featuring very massive 
O-type stars. Recently many authors have considered that the distance to $\eta$ Carinae and  
to the whole Carina region is 2.3 kpc \citep[see, e.g.,][]
{2006ApJ...644.1151S, 2011ApJS..194...14P}. There is a large discrepancy in the measured 
distances to the clusters situated within this nebula, as can be seen from Table~\ref{earlyob}. 
This large scatter in the distance occurs because, as noted by \citet{2008hsf2.book..138S}, 
the direction of the Galactic plane in the Carina nebula nearly looks down the tangent point 
of the Sagittarius-Carina spiral arm. 
The two clusters Tr~14 and Tr~16, located towards the center of the Carina nebula, have been 
extensively studied by several authors, but the debate about their distance is still open. 
\citet{1996AAS..116...75V} estimated a distance modulus of $V_0 - M_V$ = 12.5 $\pm$ 0.2 
mag for Tr~14. By applying an abnormal reddening law, \citet{2003MNRAS.339...44T} derived 
$V_0 - M_V$ = 12.1 mag. In their study, they adopted $A_V$ = 1.39$E(V - J)$ and found 
that both clusters are situated at the same distance. But in another study, \citet{2004AA...418..525C}
concluded that both clusters are situated at different distances with $V_0 - M_V$ = 12.3 $\pm$ 0.2 mag 
for Tr~14 and 13.0 $\pm$ 0.3 mag for Tr~16. Recently, \citet{2012AJ....143...41H} concluded 
that Tr~14 and Tr~16 are at the same distance within the observational errors 
($V_0 - M_V$ = 12.3 $\pm$ 0.2 mag, i.e., d = 2.9 $\pm$ 0.3 kpc). Their derived distance is 
based upon the proper motion, which is comparatively more accurate than other methods. Since 
we are concentrating on the western side of the Carina nebula containing some part of Tr~14, 
for the present study, we have adopted a distance of 2.9 kpc for CrW as given by 
\citet{2012AJ....143...41H}. 

\section{Results}

\subsection{Spectroscopically identified sources}

The MK spectral types of 15 X-ray emitting sources in the CrW region were established 
using the newly acquired spectra (see Sect.~\ref{spect}) and their comparison with 
the digital spectral classification atlas compiled by R.O. Gray and available on the
web\footnote{http://www.ned.ipac.caltech.edu/level/Gray/frames.html}. The results are
summarized in Table~\ref{xraycross}, from which we may infer that the majority of
identified sources are late-type stars (see Fig.~\ref{fig-spec} for different spectral 
types), and none of these stars features an $H\alpha$ emission.

\begin{figure*}
\centering\includegraphics[height=8cm,width=8cm,angle=0]{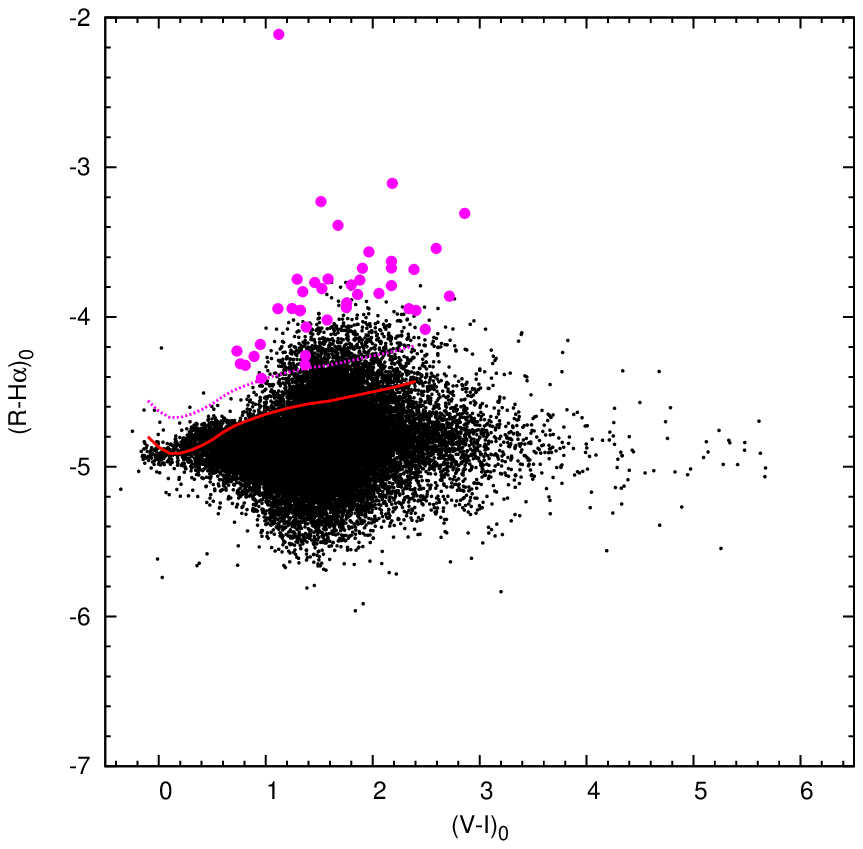}
\centering\includegraphics[height=8cm,width=8cm,angle=0]{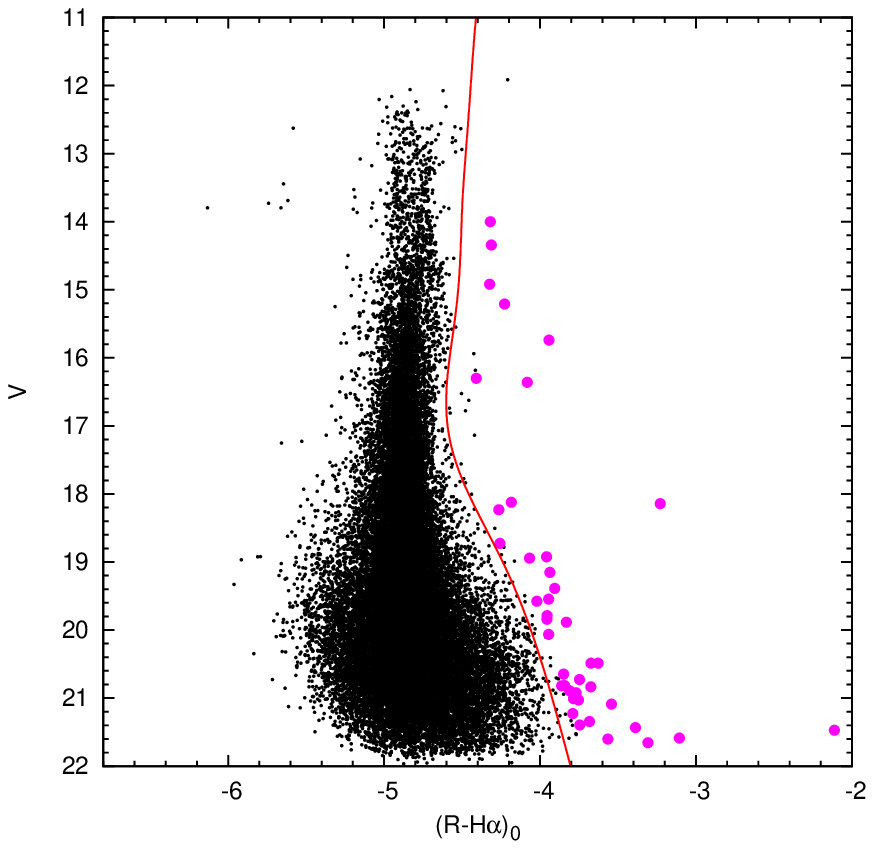}
\caption{\label{halp} Left panel: The $(R - H\alpha)_0$ index is shown as a function of the
$(V-I)_0$ color. The solid line indicates the relation for MS stars as taken from
\citet{1997AJ....114.2644S}. The dashed line (magenta) yields the thresholds for $H\alpha$
emitter candidates. Right panel: $V$ versus $(R - H\alpha)_0$ CMD. The magenta circles represent
$H\alpha$ emitter candidates. An envelope as discussed in Sect.~\ref{hapid} is indicated by a
solid line.}
\end{figure*}

Three X-ray sources (i.e. \#6, \#9, and \#20; in Table \ref{xraycross}) belong to spectral 
type O, of which \#6 and \#9 are correlated with HD~92607 and HD~92644, respectively. 
\citet{1975mcts.book.....H} classify them as O type stars (HD~92607 -- O9\,II/III and 
HD~92644 -- O9.5/B0III). Our present analysis rather favors spectral types of 
O8.5\,III for \#6 and O9.7\,V for \#9. These results broadly confirm previous 
classifications of these sources \citep[see also][]{Claeskens2011}. The X-ray properties 
of both stars are discussed in detail by \citet[][see their discussion and notes on individual 
objects]{Claeskens2011}. They find that the observed X-ray count rate for \#9 is 3.8 times 
lower than \#6, but both are quite soft. Star \#20 is identified as a reddened O7 star with 
observed $V$ = 13.03 and $(B-V)$ = 1.83 mag (cf. Table~\ref{xraycross}).    

The remaining 12 sources have counterparts that are classified as late-type stars: 
three are of spectral type A, three are F stars, and six are classified as G-type stars. 
Stars \#11, \#19, and \#37 are identified as A5 V, A1 III, and A1 V, respectively. Similarly 
\#3, \#10, and \#23 belong to F5 V, F8 V, and F3 V spectral types, respectively. 
\citet{Claeskens2011} in their study found that source \#18 is among the brightest X-ray
sources in this field; however, they could not identify any optical counterpart for this 
object from the GSC2.2 catalog. Based on IR colors, they computed the $V$ band magnitude
of this object to be in between 20.2 $-$ 21.5. Later on by visual inspection of Digital Sky 
Survey images, they found a star having brightness 18 $-$ 19 at the exact source location. 
We also found a star with magnitude 18.214 $\pm$ 0.012 (cf. Table~\ref{xraycross}, column 4) at 
a similar position. Binarity could explain why this star is brighter in the optical than expected 
from its near-IR magnitudes \citep{Claeskens2011}. It could reside in front of the Carina, but 
it could also be intrinsically brighter than a main-sequence (MS) star.
Sources \#7, \#12, \#15, \#32, \#40, and \#42 are characterized as G6 III, G9 V, G8 III, G3 V-III,
G8 V, and G9 III spectral type, respectively. Based on the observed X-ray counts, \citet{Claeskens2011}  
claim that \#42 is a variable star. It is also worthwhile to mention that two sources (\#7 and \#15) 
are identified as PMS sources (see Table~\ref{xraycross}) in the present study (cf. Sect.~\ref{x-ray}). 

\subsection{YSOs identification}\label{ysoidf}

The PMS stars (YSOs) are mainly grouped into the classes 0-I-II-III, which represent 
in-falling protostars, evolved protostars, classical T-Tauri stars (CTTSs), and weak 
line T Tauri stars (WTTSs), respectively \citep[cf.][]{1999ARAA..37..363F}. Class 0 \& I 
YSOs are so deeply buried inside the molecular clouds that they are not visible at optical 
wavelengths. The CTTSs feature disks from which the material is accreted, and emission in 
$H\alpha$ can be seen as due to this accreting material. These disks can also be probed through 
their IR excess (compared to normal stellar photospheres). WTTSs, on the contrary, have little 
or no disk material left, hence have no strong $H\alpha$ emission and IR excess. It is 
evident from the recent studies that the X-ray luminosity from WTTSs is significantly higher 
than for the CTTSs with circumstellar disks or protostars with accreting envelopes 
\citep{2004AJ....127.3537S, 2007AA...468..425T, 2008ApJ...677..401P}.
In this section we report the tentative identification of YSOs on the basis of their 
$H\alpha$ emission, IR excess, and X-ray emission. 

\subsubsection{On the basis of $H\alpha$ emission}\label{hapid}

The stars showing emission in $H\alpha$ might be considered as PMS stars or candidates, and 
the strength of the $H\alpha$ line (measured by its equivalent width `EW($H\alpha$)') is a 
direct indicator of their evolutionary stage. The conventional distinction between CTTSs and 
WTTSs is an EW($H\alpha$) $ > 10 \AA$ for the former \citep[see][]{1988cels.book.....H}. 
However, \citet{1989ARAA..27..351B} has suggested that a limiting value of 5$\AA$ might be more 
appropriate. More recently, investigators have tied the definition to the shape (width) of 
the $H\alpha$ line profile \citep[see][]{2003ApJ...582.1109W, 2003ApJ...592..282J}. In the 
study of NGC~6383, \citet{2010AA...511A..25R} find that an $H\alpha$ equivalent width of 
10$\AA$ corresponds to an $(R-H\alpha)$ index of 0.24 $\pm$ 0.04 above the MS relation of 
\citet{1997AJ....114.2644S}. They have further used this as a selection criterion for identifying 
$H\alpha$ emitters. In our study, we have considered a source as probable $H\alpha$ emitter 
only if the $(R-H\alpha)$ index is 0.24 above the MS relation by \citet{1997AJ....114.2644S}.

The $H\alpha$ filter at WFI has a special passband, therefore it cannot be directly linked to 
any existing standard photometric system \citep[see also][]{2010AA...511A..25R}. By selecting ten 
stars observed with EMMI (see Sect.~\ref{spect}), whose spectra do not exhibit $H\alpha$ emission, 
we calibrated the zero point by comparing the observed $R-H\alpha$ and dereddened $(V-I)$ with the 
$(R-H\alpha)_0$ versus the $(V-I)_0$ relation of emission free MS stars as determined by 
\citet{1997AJ....114.2644S} for NGC 2264. The $(V-I)$ color is dereddened by $E(V-I)$ value of 
$E(B-V)_{min}$ $\times$ 1.5. In Fig.~\ref{halp}, we plotted the $(R-H\alpha)_0$ vs. $(V-I)_0$ 
distribution of all the stars along with the MS given by \citet{1997AJ....114.2644S}.  

Since there is large scatter in the distribution (cf. Fig.~\ref{halp}; left panel), there may be 
false identifications of $H\alpha$ emitters. To minimize this, we introduced 
another selection criterion to identify the $H\alpha$ emitters in addition to the previous one. 
We used the $V$ vs. $(R-H\alpha)_0$ color magnitude diagram (CMD) (cf. Fig.~\ref{halp}; 
right panel) and defined an envelope that contains most of the stars following the MS. The stars 
that have a value of $(R-H\alpha)_0 - \sigma_{(R-H\alpha)}$ greater than that of the envelope of 
the MS can be assumed to be probable $H\alpha$ emitters. In our study, we therefore consider that
a star is a good $H\alpha$ emission candidate if it satisfies both conditions. We have identified
41 YSOs in our study as potential $H\alpha$ emitters, and these can be seen in Fig.~\ref{spa}. 

\subsubsection{On the basis of IR excess}\label{irex}

\begin{figure}
\centering\includegraphics[height=8cm,width=9cm,angle=0]{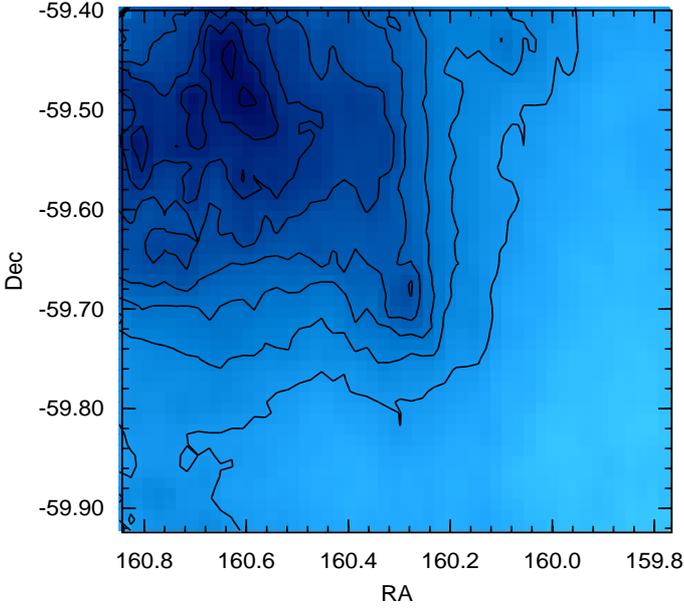}
\caption{\label{av} Column density distribution of the molecular cloud in our field of view, 
as derived from the near-infrared reddening of stars. The lowest contour corresponds to
$A_{v}$ = 3.4, the step size of the contours is 0.2.}
\end{figure}

\begin{figure*}
\centering\includegraphics[height=11cm,width=15cm,angle=0]{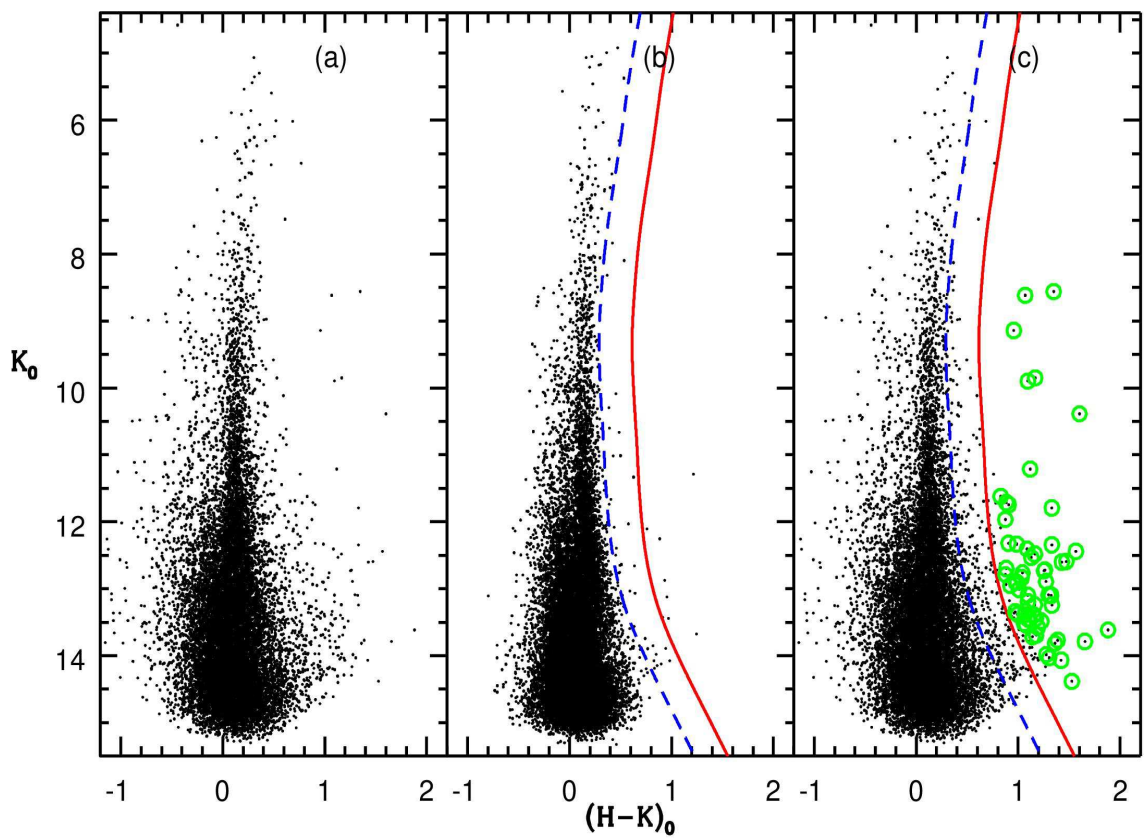}
\caption{\label{band} $K{_0}/(H-K){_0}$ CMD for (a) stars in the CrW region, (b) stars in the
field region and (c) same stars as in panel (a) along with identified probable NIR-excess stars. 
The blue dashed line represents the envelope of field CMD, whereas the red solid line demarcates the 
distribution of IR excess sources from MS stars.}
\end{figure*}

Recently \citet{2013AA...549A..67G} have obtained Herschel PACS and FIR maps that cover the 
full area of the CNC and reveal the population of deeply embedded YSOs, most of which are not 
yet visible at the MIR or NIR wavelengths. They studied the properties of the 642 objects 
that are independently detected as point-like sources in at least two of the five Herschel bands. 
For those objects that can be identified with apparently single Spitzer counterparts, they  
used radiative transfer models to derive information about the basic stellar and circumstellar 
parameters. They found that about 75\% of the Herschel-detected YSOs are Class 0 protostars and 
that their masses (estimated from the radiative transfer modeling) range from 
$\sim$1 M$_\odot$ to $\sim$10 M$_\odot$. Out of these 642 sources, 105 fall in our studied region.

Using NIR/MIR data of 2MASS and Spitzer, \citet{2011ApJS..194...14P} present a catalog of 1439 YSOs 
spanning a 1.42 deg$^2$ field surveyed by the {\it Chandra} Carina Complex Project (CCCP) \citep[for 
more details about CCCP see][]{2011ApJS..194....1T}. This field includes the major ionizing clusters 
and the most active sites of ongoing star formation within the Great Nebula in Carina. YSO candidates 
were identified via IR excess emission from dusty circumstellar disks and envelopes, using data from 
the Spitzer Space Telescope (the Vela--Carina survey) and the 2MASS database. They model the 1-24 
$\mu$m IR spectral energy distributions of the YSOs to constrain their physical properties. Their 
Pan-Carina YSO Catalog (PCYC) is dominated by intermediate-mass (2 M$_\odot$ $<$ M $\le$ 10 M$_\odot$) 
objects with disks, including Herbig Ae/Be stars and their less evolved progenitors. Out of these 
1439 sources, 136 fall in our studied region.

Recently, \citet{2011AA...530A..40P} used HAWK-I at the ESO VLT to produce a deep and wide NIR 
survey that is deep enough to detect the full low-mass stellar population (i.e. down to 
$\sim$0.1M$_\odot$ and for extinctions up to $A_V \sim$15 mag) in all the important parts of the 
CNC, including the clusters Tr 14, 15, and 16, as well as the South Pillars region. They analyzed 
CMDs to derive information about the ages and masses of the low-mass stars. Unfortunately, their 
surveyed region does not cover our studied region. Therefore for the present study we used 
NIR data from the 2MASS survey to identify sources with IR excess. We used the following 
scheme to make the distinction between the sources with IR excess and those that are simply reddened 
by dust along the line of sight \citep{2005ApJ...632..397G}. First we measure the line of sight 
extinction to each source as parameterized by the $E_{H-K}$ color excess due to the dust present 
along the lines of sight. For objects where we have $J$ photometry in addition to $H$ and $K_s$ 
with the condition that they are positioned above the extension of CTTSs locus and have color 
$[J-H] \ge 0.6$, we used the equations given by \citet{2009ApJS..184...18G} to derive the 
adopted intrinsic colors. The difference between the intrinsic color and the observed one will 
give the extinction value. Once we had the extinction value for the stars, we generated an 
extinction map for the whole CrW region. The extinction values in a sky plane were calculated 
with a resolution of 5 arcsec by taking the mean of extinction value of stars in a box having a 
size of 17 arcsec. The resulting extinction map, smoothed to a resolution of 0.6 arcmin, is shown 
in Fig.~\ref{av}. This IR extinction map represents the column density distribution of the molecular
cloud associated with the CrW region. We can clearly see the high density region toward the 
northeast of CrW and then the density following the dust lane as visible in the 4.6~$\mu$m image 
(cf. Sect.~\ref{diss}, see Fig.~\ref{spa}).
Thanks to less extinction, longer wavelength observations can penetrate deeper inside the 
nebulosity than do shorter wavelengths. For this reason, there are many stars in the CrW region
that do not have $J$ band photometry. Once we constructed the extinction map, we used this 
to also deredden the stars having no $J$ band detection. 
Here it is worthwhile to note that we used CTTSs loci to estimate reddening by back-tracing 
all the stars located above the CTTSs loci or its extension to the CTTSs loci or its extension. 
It is quite probable that the genuine CTTSs are mixed with deeply embedded MS stars that 
could fall above the CTTSs locus in the 2MASS color-color diagram.
Therefore, when we deredden this mixed sample of stars, the 
reddening value for CTTSs will get overestimated because of their surrounding cocoon of dust/gas, 
whereas for the MS stars, it will get underestimated because the intrinsic color of MS stars lies 
below CTTSs loci.

In Fig.~\ref{band}, we have plotted the dereddened NIR CMDs, $K_0$ versus $(H - K)_0$, for the CrW 
region and the nearby reference field region having same area as CrW. Since both the field and CrW 
region CMDs are dereddened by the same technique, the underestimation/overestimation of the $A_v$ value 
will not affect our analysis much. However, owing to the clumpy nature of the molecular cloud associated 
with the CrW region, the dereddened CMD for CrW will show more scatter than the field CMD. A comparison 
with the reference field CMD reveals that there might be many stars showing excess emission that is 
apparent from their distribution at $(H-K)_0 \lesssim 0.6$ mag. Therefore, we defined an envelope 
representing a cut-off line (Figs.~\ref{band}b,c) on the basis of the CMD of the CrW 
region and of the field one. We then designed an additional envelope (solid line in red) shifted to 
the red from the preceding one by an amount corresponding to $A_v$ = 5 (to compensate for the scattering 
due to the clumpy nature of molecular clouds). Doing that, we aim at isolating probable NIR excess 
stars from reddened MS stars.
Since we know that the photometric error is larger at the fainter end of the CMD, the shape of 
the cut-off line at the fainter end is adjusted accordingly. 
All the stars having a color 
`$(H-K)_0 - \sigma_{(H-K)_0}$' greater than the red cut-off line might have an 
excess emission in the $K$ band and thus can reasonably be considered to be probable YSOs 
\citep[see also][]{2012ApJ...759...48M}. While this sample is dominated by YSOs, it may also contain 
the following types of contaminations: variable stars, dusty asymptotic giant branch (AGB) stars, 
unresolved planetary nebulae, and background galaxies \citep{2008AJ....136.2413R, 2011ApJS..194...14P}. 
We used the CMD of the reference field covering the same area as the CrW region to calculate 
the fraction of contaminating objects in our sample. The reference field was around 1.5$^{\circ}$  
westward from the center of the CrW (cf. Fig.~\ref{totalregion}).

The number of probable NIR 
excess stars in the reference field is about 8 whereas the number of probable NIR excess stars 
in the CrW region is 60. This means that we have a contamination of about 13\% in our sample. 
The majority of these probable NIR excess stars follow the high density region in the CrW 
region (see Fig.~\ref{spa}) and may be deeply embedded in that nebulosity. \citet{2011ApJS..194...14P} 
have identified many YSOs that are mainly in the irradiated surface of the cloud (see Fig.~\ref{spa}) 
but recently, based on Herschel FIR data, \citet{2013AA...549A..67G} have identified YSOs that are 
also located within the dark lane of the CrW region.

\begin{figure}
\centering\includegraphics[height=8.5cm,width=8.5cm,angle=0]{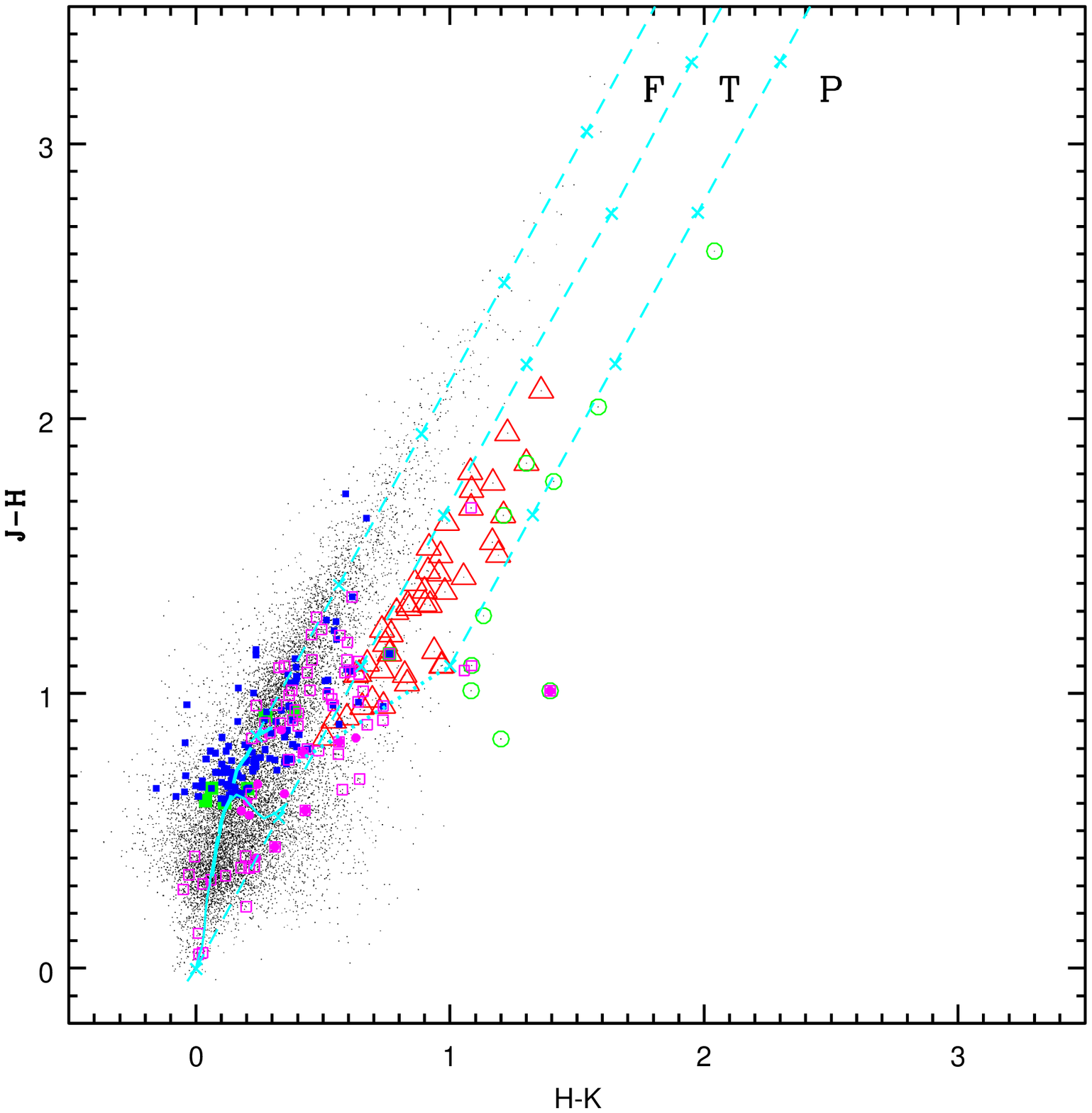}
\caption{\label{nir-yso} $(J - H)/(H - K)$ color-color diagram of sources detected
in the $JHKs$ bands in the CrW region. The sequences of dwarfs (solid curve) and giants
(thick dashed curve) are from \citet{1988PASP..100.1134B}. The dotted line represents the
locus of T Tauri stars \citep{1997AJ....114..288M}. Parallel dashed straight lines 
represent the reddening vectors \citep{1981ApJ...249..481C}. The crosses on the dashed 
lines are separated by $A_V$ = 5 mag. YSOs are also shown. Open magenta squares = Spitzer; 
filled magenta circles = $H\alpha$; filled squares = X-ray emitting WTTSs 
(green = {\it XMM-Newton}, blue  = {\it Chandra}); open red triangles = CTTSs and
open green circles = probable NIR-excess sources (see text for the classification 
scheme).
}
\end{figure}

YSOs such as CTTSs, WTTSs, and Herbig Ae/Be stars tend to occupy different 
regions on the NIR TCDs. In Fig.~\ref{nir-yso}, we have 
plotted the NIR TCD using the 2MASS data for all the sources lying in the observed region. 
All the 2MASS magnitudes and colors were converted into the California Institute of 
Technology (CIT) system\footnote{\url{http://www.astro.caltech.edu/~jmc/2mass/v3/transformations/}}.
All the curves and lines are also in the CIT system. 
The shown reddening vectors are drawn from the tip 
(spectral type M4) of the giant branch (``upper reddening line"), from the base 
(spectral type A0) of the MS branch (``middle reddening line") and from the tip of the 
intrinsic CTTSs line (``lower reddening line"). The extinction ratios $A_J/A_V = 0.265, 
A_H/A_V = 0.155$ and $A_K/A_V=0.090$ have been taken from \citet{1981ApJ...249..481C}.  
We classified the sources according to three regions in this diagram \citep[cf.][]{2004ApJ...608..797O}. 
The `F' sources are located between the upper and middle reddening lines and are considered 
to be either field stars (MS stars, giants) or Class III and Class II sources with small 
NIR-excess. `T' sources are located between the middle and lower reddening lines. These sources 
are considered to be mostly CTTSs (or Class II objects) with large NIR-excess. There may be an 
overlap of Herbig Ae/Be stars in the `T' region \citep{1992ApJ...397..613H}. `P' sources are 
those located in the region redward of the lower reddening line and are most likely Class 0/I 
objects (protostellar-like objects; \citet{2004ApJ...608..797O}). It is worthwhile also mentioning 
that \citet{2006ApJS..167..256R} show that there is a significant overlap between 
protostars and CTTSs. The NIR TCD of the observed region (Fig.~\ref{nir-yso}) indicates that a 
significant number of sources that have previously been identified as probable NIR-excess stars 
lie in the `T' region. Forty-one sources have been designated as CTTSs in our study with 
the condition that they fall in the `T' region of the NIR TCD (Fig.~\ref{nir-yso}) and are redward 
of the dashed blue cut-off line in the dereddened CMD (Fig.~\ref{band}). 
We also have plotted probable NIR excess sources 
that are detected in `$J$' band (10 out of 60). Most of these sources are located in the `P' 
region in Fig.~\ref{nir-yso}, which means that they are most likely Class 0/I objects.

\subsubsection{On the basis of X-ray emission}\label{x-ray}

The NIR-excess-selected YSO candidate samples are generally considered incomplete because 
the NIR-excess emission in young stars disappears on timescales of just a few Myr 
\citep[see][]{2007prpl.conf..345B}. At an age of $\sim$3 Myr, only $\sim$50\% of the young 
stars still show NIR excesses, and by $\sim$5 Myr this fraction is reduced to $\sim$15\% 
\citep{2011AA...530A..34P}. Since the expected ages of most young stars in the CNC are several 
Myr, any IR-excess-selected YSO sample will be highly incomplete. To tackle this problem, 
we used the X-ray emitting point sources in the region to identify YSO candidates. 
The X-ray detection methods are sensitive to young stars that have already dispersed their 
circumstellar disks, thus avoiding the bias introduced when selecting samples only based 
on IR excess \citep{2011AA...530A..34P}.

\begin{figure}
\centering\includegraphics[height=8cm,width=9cm,angle=0]{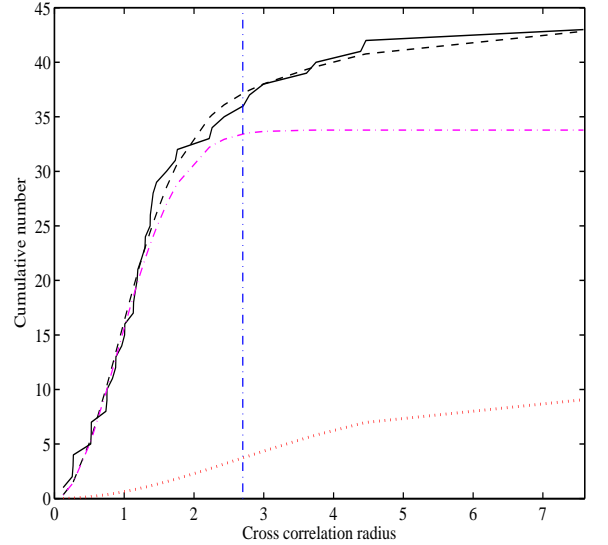}
\caption{Cumulative numbers of correlations between the X-ray detected sources and the WFI 
catalog. The thick curve represents the observed numbers, the dashed curve shows the best 
fit, and the dot-dashed line (magenta) and dotted (red) curves correspond to the numbers of 
real and spurious sources, respectively. The vertical line indicates the optimal correlation 
radius r$_c$.}
\label{xo}
\end{figure}

\subsubsection*{{\it XMM-Newton observations}}

The {\it XMM-Newton} satellite has observed the CrW region in the course of the study of the 
massive binary WR~22. The corresponding results have been presented in separate papers 
\citep[see][]{2009AA...508..805G, Claeskens2011}. In this section we cross-correlated the 
sources detected in our photometric data of the CrW region with the positions of 43 X-ray 
sources from \citet{Claeskens2011}. The positions of the X-ray sources as given by 
\citet[][columns 8 and 9 of their table 2]{Claeskens2011} refer to the astrometric frame
as determined from the {\it{XMM-Newton}} on-board Attitude and Orbit Control System. The 
cross-correlation with the GSC and the present optical catalog suggests making a small 
correction. We therefore suggest decreasing the right ascension of \citet{Claeskens2011} 
by 0$\farcs$25 and increasing the declination by 0$\farcs$96. No rotation was detected. 
We adopt these new positions for the 43 X-ray sources.

We defined an optimal cross-correlation radius to find a compromise between correlations 
missed due to astrometric errors and spurious associations in the CrW field. To derive the 
optimal correlation radius, we applied the technique of \citet{Jeffries1997}. In this method, 
the distribution of the cumulative number of cataloged sources as a function of the
cross-correlation radius $r_c$ is given by

\begin{eqnarray*}
\Phi(d \leq r_c) & = & A\,\left[1 - \exp{\left(\frac{-r_c^2}{2\,\sigma^2}\right)}\right] \\
& + & (N - A)\,\left[1 - \exp(-\pi\,B\,r_c^2)\right].
\end{eqnarray*}

In this equation $N$, $A$, $\sigma$, and $B$ represent the total number of cross-correlated X-ray 
sources ($N = 43$), the number of true correlations, the uncertainty in the X-ray source position, 
and the surface density of the catalog of photometric sources, respectively. 
In the course of fitting the integrated number of correlations with the CrW photometric catalog
as a function of the separation (where 43 X-ray sources have an optical source located closer than 
8 arcsec), we derived the fitting parameters as $A = 33.78$, $\sigma = 0.9$\,arcsec, and 
$B = 2.3 \times 10^{-2}$\,arcsec$^{-2}$ (see Fig. \ref{xo}). The optimal correlation radius was chosen 
to be $r_c$ = 2.7$\,\arcsec$, which implies that there is no more than one spurious association among 
the 34 correlations; i.e., the optical counterparts of 79\% of the {\it XMM-Newton} sources in the CrW 
field should thus be reliably identified. 

\citet{Claeskens2011} also cross-correlated these X-ray sources ($N = 43$), but they searched the optical 
counterparts using the Guide Star Catalog version 2.2 (GSC2.2). Their fitting parameters are $A = 35.4$,
$B = 2 \times 10^{-3}$\,arcsec$^{-2}$, and $\sigma = 1.8$\,arcsec. The number of true correlations ($A$) 
of both studies are very much consistent. The surface density ($B$) of the catalog of optical sources 
in the present study is an order of magnitude higher than in \citet{Claeskens2011}, while our $\sigma$ 
value is half that of these authors. The results of the cross-identification are listed in 
Table~\ref{xraycross}. When several optical counterparts are present, only the closest one is given. 
The first column is the ID of the X-ray sources from \citet{Claeskens2011}. Columns 2 and 3 are their 
respective RA and Dec in degree (from our photometric catalog). In the next columns the $V$ magnitude, 
colors $(V-I)$, $(V-R)$, $(B-V)$ and $(U-B)$ are reported. Sources with a spectral classification are 
mentioned in the next-to-last column.  

\begin{table*}
\centering
\scriptsize
\caption{Cross-identification of 43 X-ray sources from \citet{Claeskens2011} with CrW optical photometry.
Stars brighter than $V$ = 11.3 are from the literature. The YSOs identified in Section~\ref{ysoidf} are 
also mentioned in the last column.}
\label{xraycross}
\begin{tabular}{cccccccccc}
\hline
ID(X-ray)&$\alpha$(J2000)&$\delta$(J2000)& $V$ & $(V-I)$ & $(V-R)$ & $(B-V)$ & $(U-B)$ & Spectral type & YSO number$^{\dagger}$ \\ 
\hline
\#1 &   159.894496 &-59.737510 & 20.386 & 2.860  &  1.438 &  2.250 & N/A    & --         &183 \\         
\#2 &   159.947016 &-59.608981 & 12.991 & 0.809  &  0.436 &  0.653 & -0.042 & --         &--  \\         
\#3 &   159.948029 &-59.746218 & 10.470 & N/A    &  N/A   &  0.440 & N/A    & F5\,V      &--  \\         
\#4 &   159.983333 &-59.618056 & 11.279 & N/A    &  N/A   &  0.668 & N/A    & --         &--  \\         
\#5 &   160.042973 &-59.619011 & 17.720 &  2.299 &  1.195 &  1.721 &  1.051 & --         &57  \\         
\#6 &   160.051779 &-59.802809 &  8.140 & N/A    &  N/A   & -0.020 & -0.780 & O8.5\,III  &--  \\         
\#7 &   160.069977 &-59.534346 & 14.720 &  1.386 &  0.787 &  1.006 &  0.673 & G6\,III    &13 \\        
\#8 &     --       &  --       &  --    &   --   &   --   &    --  &  --    & --         &--  \\         
\#9 &   160.132138 &-59.778854 &  8.880 & N/A    & N/A    & -0.030 & -0.900 & O9.7\,V    &--  \\         
\#10 &  160.161981 &-59.462519 & 12.839 &  0.807 &  0.419 &  0.590 &  0.028 & F8\,V      &--  \\         
\#11 &  160.174252 &-59.621872 & 12.609 & 0.429  & 0.258  &  0.320 &  0.108 & A5\,V      &--  \\   
\#12 &  160.174412 &-59.539547 & 14.974 & 0.954  & 0.540  &  0.736 &  0.149 & G9\,V      &--  \\ 
\#13 &  160.184003 &-59.826372 & 17.565 &  3.222 &  1.551 &  2.103 &  1.018 &  --        &--  \\         
\#14 &  160.193248 &-59.700809 & 19.680 &  2.144 &  1.115 &  1.544 & N/A &  --        &--  \\         
\#15 &  160.214424 &-59.622191 & 15.365 &  1.332 &  0.716 &  1.109 &  0.609 & G8\,III    &18  \\         
\#16 &    --       &  --       &  --    &   --   &   --   &    --  &  --    & --         &--  \\         
\#17 &  160.229239 &-59.639602 & 17.422 &  1.552 &  0.835 &  1.217 &  0.607 & --         &48  \\         
\#18 &  160.230446 &-59.710999 & 18.214 &  1.586 &  0.928 &  1.086 &  0.373 & F8\,V      &--  \\         
\#19 &  160.235367 &-59.862561 & 11.363 &  0.340 & N/A    & N/A    & N/A    & A1\,III    &--  \\         
\#20 &  160.247070 &-59.457005 & 13.031 &  1.825 &  0.860 &  1.340 & -0.031 &  O7         &--  \\
\#21 &    --       &  --       &  --    &   --   &   --   &    --  &  --    & --         &--  \\         
\#22$^{a}$ & 160.322983 &-59.676915 &   6.420    & N/A    & N/A    &  0.080 &  -0.730 & --    &--  \\    
\#23 &  160.335850 &-59.589894 & 11.692 &  0.718 & N/A    &  0.841 & -0.285 & F3\,V      &--  \\         
\#24 &  160.338038 &-59.659587 & 17.131 &  1.316 &  0.758 &  0.978 &  0.259 & --         &--  \\         
\#25 &    --       &  --       &  --    &   --   &   --   &    --  &  --    & --         &--  \\         
\#26 &  160.362639 &-59.656561 & 16.698 &  1.334 &  0.774 &  0.896 &  0.494 & --         &--  \\         
\#27 &  160.364564 &-59.686933 & 16.339 &  1.174 &  0.658 &  0.903 &  0.215 & --         &--  \\         
\#28 &  160.387108 &-59.604288 & 19.416 &  2.737 &  1.136 &  1.667 & N/A    & --         &--  \\         
\#29 &  160.426017 &-59.635431 & 16.982 &  1.335 &  0.712 &  1.054 &  0.482 & --         &--  \\         
\#30 &  160.437963 &-59.807270 & 19.050 &  2.058 &  0.922 &  1.818 & N/A    & --         &--  \\         
\#31 &  160.464076 &-59.720771 & 12.626 &  0.782 &  0.442 &  0.550 & -0.019 & --         &--  \\  
\#32 &  160.478811 &-59.689836 & 13.146 &  1.224 &  0.793 &  0.935 &  0.923 & G3\,V-III  &--  \\         
\#33 &    --       &  --       &  --    &   --   &   --   &    --  &  --    & --         &--  \\         
\#34 &    --       &  --       &  --    &   --   &   --   &    --  &  --    & --         &--  \\         
\#35 &    --       &  --       &  --    &   --   &   --   &    --  &  --    & --         &--  \\         
\#36 &    --       &  --       &  --    &   --   &   --   &    --  &  --    & --         &--  \\         
\#37 &  160.523400 &-59.604232 & 14.022 &  0.554 &  0.316 &  0.357 &  0.158 & A1\,V      &--  \\         
\#38 &  160.556738 &-59.599388 & 18.204 &  2.910 &  1.446 &  2.081 &  0.479 & --         &87  \\         
\#39 &  160.564481 &-59.565945 & 19.215 &  2.738 &  1.145 &  1.679 & N/A    & --         &--  \\         
\#40 &  160.603836 &-59.669004 & 14.842 &  1.105 &  0.636 &  0.576 &  0.164 & G8\, V     &--  \\         
\#41 &  160.636112 &-59.611374 & 16.242 &  1.834 &  0.975 &  1.461 &  1.079 & --         &30  \\         
\#42 &  160.652591 &-59.731771 & 15.372 &  1.454 &  0.764 &  1.096 &  0.746 & G9\, III   &--  \\         
\#43 &  160.679111 &-59.591296 & 15.898 &  1.365 &  0.738 &  1.165 &  0.425 & --         &--  \\         

\hline
\end{tabular} \\
$^{a}$ WR~22 itself;
$^{\dagger}$The YSO entry numbers are from Table~\ref{sammpleyso}.
\end{table*}

\subsubsection*{{\it Chandra} X-ray observations}\label{cxo}
Recently, a wide area (1.42 deg$^2$) of the Carina complex has been mapped by the {\it Chandra} 
X-ray Observatory \citep[CCCP,][]{2011ApJS..194....1T}. These images were obtained with the 
Advanced CCD Imaging Spectrometer \citep[ACIS;][]{2003SPIE.4851...28G}. This CCCP study 
mainly includes the 
data from the ACIS-I array, although ACIS-S array CCDs S2 and S3 were also operational during 
the observations. But most of the sources on S2 and S3 are crowded and dominated by the background 
in the CCCP data \citep{2011ApJS..194....1T}. In this survey, 14369 X-ray sources were detected 
over the whole CCCP survey region. Out of these, the CrW region contains 1465 sources. Since the 
on-axis {\it Chandra} PSF is 0.5$\,\arcsec$ and because it degrades at 
large off-axis angles \citep[see, e.g.,][]{2005ApJS..160..319G, 2010ApJ...714.1582B}, we have 
taken an optimal matching radius of 1 arcsec to determine the optical/NIR counterparts of these 
X-ray sources. This size of the matching radius is well established in other studies as well 
\citep[see, e.g.,][]{2002ApJ...574..258F, 2007ApJS..168..100W}. We identified 469 sources that 
have 2MASS NIR counterparts and fall in the CrW region.

\subsubsection*{Classification of X-ray emitters based on NIR TCD}

We have identified WTTSs based on their X-ray emission, as well as on their respective
position in NIR TCD (Fig.~\ref{nir-yso}) through {\it Chandra} and {\it XMM-Newton} observations.
The sources having X-ray emission and lying in the `F' region above the
extension of the intrinsic CTTSs locus, as well as sources having $(J-H) \geq 0.6$ mag and
lying to the left of the first (leftmost) reddening vector (shown in Fig.~\ref{nir-yso}) 
are assigned as WTTSs/Class III sources 
\citep[see, e.g.,][]{2008MNRAS.384.1675J, 2008MNRAS.383.1241P,
2012PASJ...64..107S}. Here it is worthwhile to mention that some of the X-ray sources classified as
WTTSs/Class III sources, lying near the middle reddening vector, could be CTTSs/ Class II sources.
Out of 34 ({\it XMM-Newton}) and 469 ({\it Chandra}) sources, 7 and 119, respectively, were identified
as WTTSs, with 4 in common.
These are identified in Table~\ref{sammpleyso} (by numbers 4-5 in the last column and filled
squares in Fig.~\ref{nir-yso}).

\subsection{Age and mass of YSOs}\label{ageandmass} 

\subsubsection{Using NIR CMD}\label{nircmd}

\begin{figure}
\centering\includegraphics[height=8.5cm,width=8.5cm,angle=0]{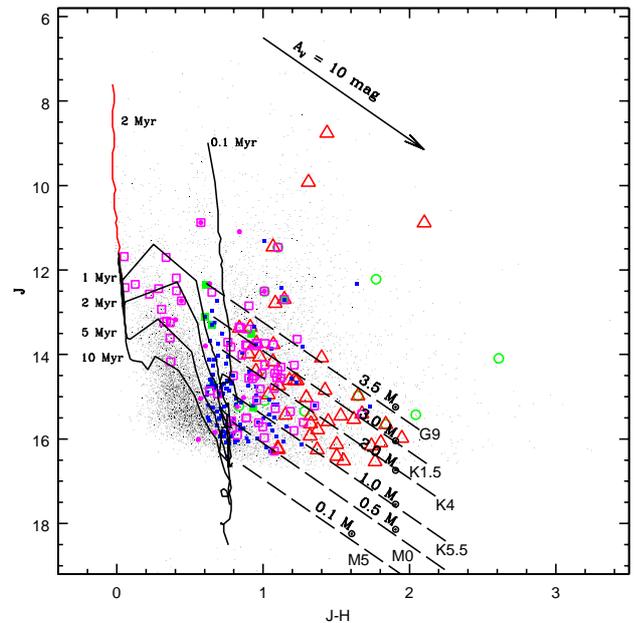}
\caption{ \label{nir-yso1}{$J/(J - H)$ CMD for the stars in the CrW region. The isochrone of 2 
Myr (Z = 0.02) by \citet{2008AA...482..883M} and PMS isochrones of age 0.1, 1, 2, 5 and 
10 Myr taken from \citet{2000AA...358..593S} corrected for a distance of 2.9 kpc and a 
reddening $E(B - V)_{min}$ = 0.25 are also shown. The symbols are the same as in 
Fig.~\ref{nir-yso} (see Sect.~\ref{nircmd} for the classification scheme). The shown masses 
and spectral types have been taken from the 1 Myr PMS isochrone of \citet{2000AA...358..593S}.}}
\end{figure}

The CMDs are useful tools for studying the nature of the stellar population 
within star-forming regions. In Fig.~\ref{nir-yso1}, we plotted the $J/(J - H)$ CMD 
for all the YSOs identified in the previous sections having NIR counterparts and located in 
the CrW region. For cross-matching the $H\alpha$, X-ray, Spitzer, and Herschel identified 
YSOs with the 2MASS data, we took a search radius of 1 arcsec. For the FIR Herschel 
identified YSOs, we did not find any NIR counterpart. We used the relation 
$A_{J} /A_{V}$ = 0.265, $A_{H} /A_{V}$ = 0.155 \citep{1981ApJ...249..481C}, isochrones 
for age- 2 Myr and PMS isochrones for ages 0.1, 1, 2, 5, and 10 Myr by \citet{2008AA...482..883M} 
and \citet{2000AA...358..593S}, respectively, to plot the CMD assuming a distance of 2.9 kpc 
and an extinction $E(B - V)_{min}$ = 0.25. In the present analysis, we used $R_V$ = 3.7 
as discussed in Sect.~\ref{rdl}. Different classes of probable YSOs are also shown in the figure. 
Most of the probable T Tauri, $H\alpha$ emission stars and IR excess stars have an apparent age 
under 1 Myr. The Spitzer identified YSOs are located mainly in two groups, one shows ages 
less than 1 Myr, whereas other groups have ages between 1$-$10 Myr. \citet{2010MNRAS.406..952S} 
also find that the majority of YSOs in Carina have ages of $\sim$1 Myr.

\begin{figure}
\centering\includegraphics[height=8cm,width=8cm,angle=0]{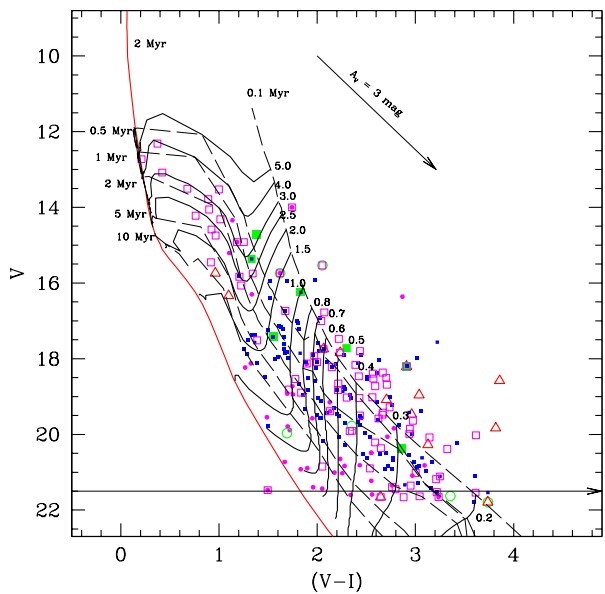}
\caption{\label{cmd_yso} $V/(V-I)$ CMD for all the detected YSOs (symbols as in Fig.~\ref{nir-yso},
see Sect.~\ref{irex} for details). The isochrone
for 2 Myr by \citet{2008AA...482..883M} (continuous line) and PMS isochrones for 1, 2, 5, and 10 Myr
by \citet{2000AA...358..593S} (dashed lines) are also shown. All the isochrones are corrected
for a distance of 2.9 kpc and reddening $E(B-V)=0.25$. The horizontal line with arrow is the
completeness limit of the observations.}
\end{figure}

The mass of the probable YSO candidates can be estimated by comparing their location on 
the CMD with the evolutionary models of PMS stars. The slanted dashed curve, taken from 
\citet{2000AA...358..593S}, denotes the locus of 1 Myr old PMS stars having masses 
in the range of 0.1 to 3.5 $M_\odot$. To estimate the stellar masses, the $J$ luminosity 
is recommended rather than that of $H$ or $K$, because the $J$ band is less affected by the 
emission from circumstellar material \citep{1988ApJ...330..350B}. The majority of the YSOs 
have masses in the range 3.5 to 0.5 $M_\odot$, indicating that these may be T Tauri stars. 
A few stars with a mass higher than 3.5 $M_\odot$ may be candidates for Herbig Ae/Be stars.
\citet{2013AA...549A..67G} state that this region exhibits a low number of very 
massive stars. However, since the more massive stars form more quickly and tend to be more 
obscured, and since they may not exhibit the same signatures of youth for as long a time as 
lower mass stars, a more extensive analysis is required to confirm their presence or absence
in this region.

The NIR counterparts of YSOs in a nebular star-forming 
region are more easy to find than their optical counterparts. Therefore in the NIR CMD we 
have statistically more YSOs but to derive the exact age/mass of individual YSOs is somewhat 
difficult since at the lower end of the NIR CMD, the isochrones of different ages and masses 
nearly coincide with each other. Age and mass of individual YSOs can be derived more accurately 
using the optical CMDs.

\begin{table*}
\centering
\scriptsize
\caption{Sample of the optically identified YSOs along with their derived ages and masses.
Error bars in magnitude and color represent formal internal (comparative) errors and do not 
include the color transformation and zero-point uncertainties.}
\label{sammpleyso}
\begin{tabular}{cccccccc} \hline 
         ID &  $\alpha$(J2000)  & $\delta$(J2000) & $V\pm \sigma$ & $(V-I)\pm \sigma$ & Age $\pm \sigma$ & Mass $\pm \sigma$& Technique \\
            & ($\degr$)         & ($\degr$)       & (mag)         &  (mag)            & (Mys)            & (M$_\odot$)      & $^{1,2,3,4,5,6}$\\ 
\hline
(1) &    (2)    &    (3)     &         (4)       &        (5)       &      (6)      &     (7)      &  (8)  \\ \hline
1   & 160.544858& -59.643538 &  12.316$\pm$0.009 &  0.373$\pm$0.018 &  0.9$\pm$0.2  & 3.7$\pm$0.2  &   1   \\
2   & 160.586232& -59.898926 &  12.726$\pm$0.011 &  0.211$\pm$0.018 &  2.6$\pm$2.1  & 4.8$\pm$0.3  &   1   \\
3   & 160.556561& -59.735036 &  13.079$\pm$0.011 &  0.422$\pm$0.014 &  1.4$\pm$0.2  & 2.8$\pm$0.3  &   1   \\
4   & 159.827622& -59.759030 &  13.508$\pm$0.006 &  0.677$\pm$0.010 &  2.5$\pm$0.4  & 2.0$\pm$0.3  &   1   \\
5   & 160.509158& -59.674841 &  13.527$\pm$0.009 &  1.000$\pm$0.014 &  1.4$\pm$0.2  & 3.4$\pm$0.3  &   1   \\
--  &    --     &    --      &          --       &         --       &       --      &      --      &   --  \\ 
\hline
\end{tabular}\\ 
$^1$ Spitzer identified sources, $^2$ $H\alpha$ sources, $^3$ CTTS, $^4$ {\it Chandra} sources,
$^5$ {\it XMM-Newton} sources, $^6$ Probable NIR excess \\
(This table is available in its entirety in a machine-readable form in the online journal. A portion is shown here for 
guidance regarding its form and content.)
\end{table*}

\begin{figure*}
\centering\includegraphics[height=5.5cm,width=6.8cm,angle=0]{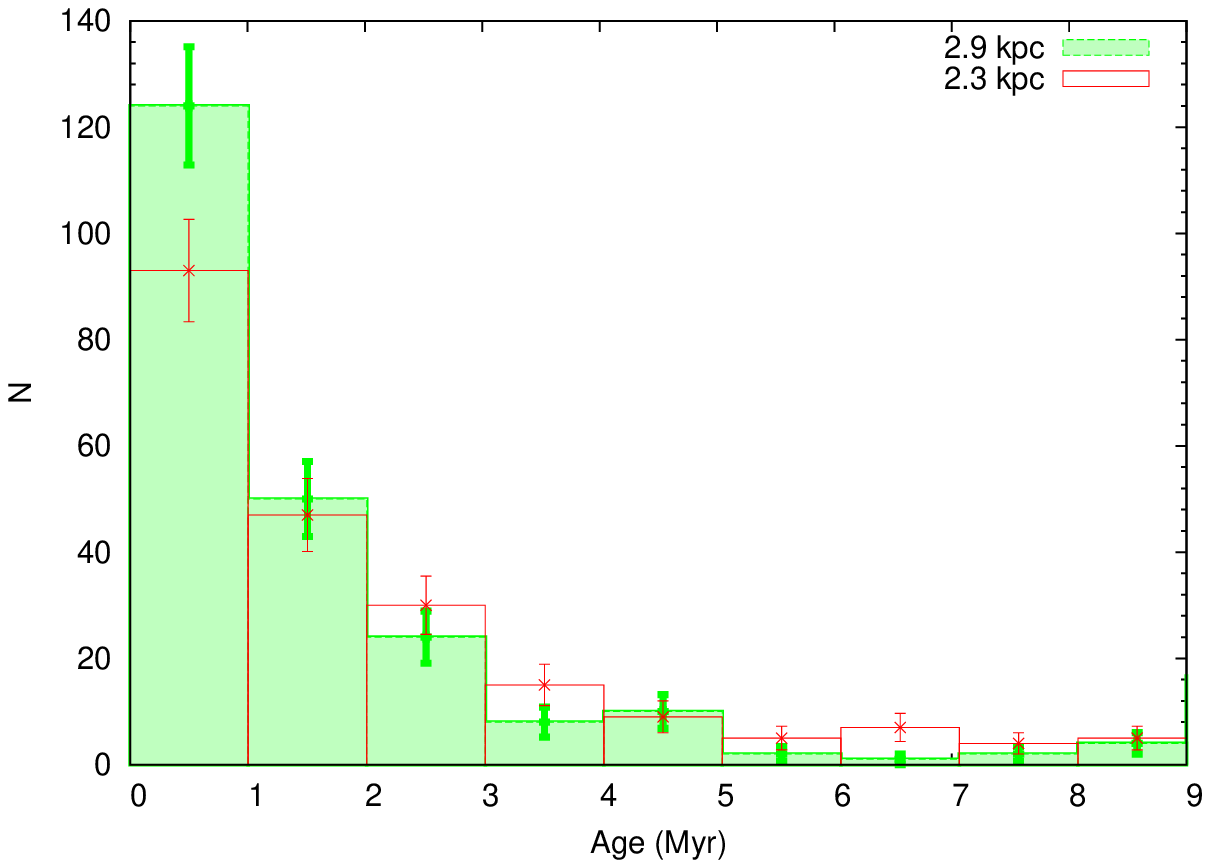}
\centering\includegraphics[height=5.5cm,width=6.8cm,angle=0]{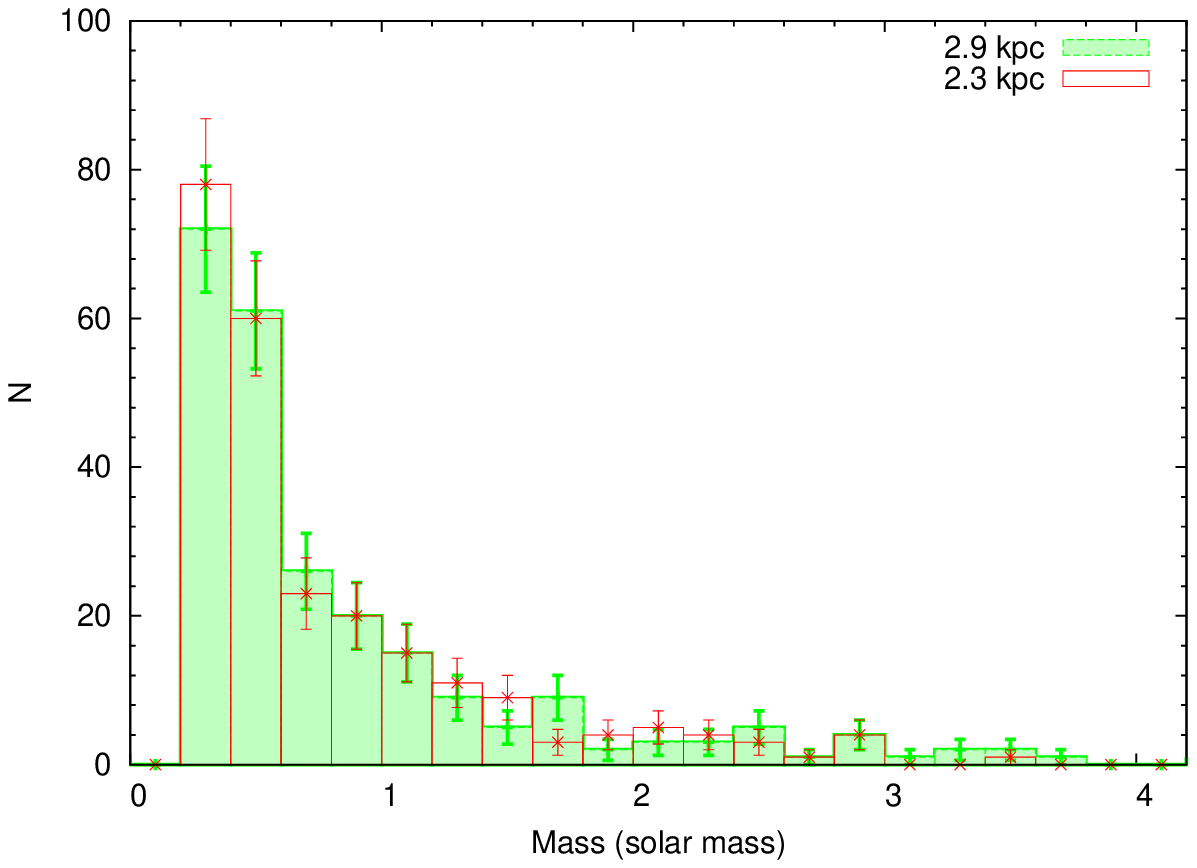}
\caption{\label{hist} Histograms showing the distribution of YSOs ages (left panel) and
masses (right panel) in the observed region.
The green and red histograms are for the estimated ages and masses of
YSOs assuming a distance of 2.9 kpc and 2.3 kpc, respectively.
The error bars along ordinates represent $\pm\sqrt N$ Poisson errors.
}
\end{figure*}

\subsubsection{Using optical CMD}\label{optcmd}
In Fig.~\ref{cmd_yso}, the $V /(V - I)$ CMD has been plotted for the optical counterparts of 
YSOs identified in Sect.~\ref{ysoidf}. We have taken the same 1 arcsec matching radius for identifying 
the optical counterparts of the presumed YSOs. Here also we have not found any optical counterpart 
of the Herschel identified YSOs. The dashed lines (for different ages 0.1, 0.5, 1, 2, 5 and 10 Myr) 
show PMS isochrones by \citet{2000AA...358..593S} and the post-main-sequence isochrone
(continuous line) for 2 Myr by \citet{2008AA...482..883M}. These isochrones are corrected for 
the CrW distance (2.9 kpc) and minimum reddening ($E(B - V ) = 0.25$ mag, see previous section). 
It is clear from Fig.~\ref{cmd_yso} that a majority of the sources have ages $<$ 1 Myr with a 
possible age spread up to 10 Myr. 

The age and mass of the YSOs have been derived using the $V /(V - I)$ CMD. 
The \citet{2000AA...358..593S} isochrones have very coarse resolution (30 points
for their whole mass range of 0.1 to 7 M$_\odot$); therefore, for a better estimation of mass,  
these isochrones were interpolated (2000 points). We used photometric errors along with 
the error in the distance modulus and reddening to draw an error box around each data point. 
In this box, we generated 500 random points using Monte Carlo simulations. For each 
generated point, we calculated the age and mass as derived from the nearest passing isochrone. 
For this study we used a bin size of 0.1 Myr for the \citet{2000AA...358..593S} isochrones. 
At the end we took the mean and standard deviation as the final derived values.

It is important to note that estimating the ages and masses of the PMS stars by 
comparing their locations in the CMDs with theoretical isochrones is prone to both random  
and systematic errors \citep[see][]{2005astro.ph.11083H, 2008ASPC..384..200H, 
2009MNRAS.396..964C, 2011MNRAS.415.1202C}. The effect of random errors due to photometric 
errors and reddening estimation in determining the ages and masses has been 
evaluated by propagating the random errors to their observed measurements by assuming a 
normal error distribution and using Monte Carlo simulations \citep[cf.][]{2009MNRAS.396..964C}. 
The systematic errors could be due to the use of different PMS evolutionary models and an 
error in the distance estimation. \citet{2011MNRAS.415..103B} mention that the ages may 
be incorrect by a factor of two owing to systematic errors in the model.
The presence of variable extinction in the region will not affect the age estimation 
significantly because the reddening vector in the $V /(V - I )$ CMD is nearly parallel to the PMS 
isochrone.

The presence of binaries may also introduce errors into the age determination. Binarity will 
brighten the star, consequently the CMD will yield a lower age estimate. In the case of an 
equal-mass binary, we expect an error of $\sim$50\% to $\sim$60\% in the PMS age estimation. 
However, it is difficult to estimate the influence of binaries/variables on the mean age 
estimation because the fraction of binaries/variables is not known. In the study of TTSs 
in the H{\sc ii} region IC~1396, \citet{2011MNRAS.415..103B} point out that the number of 
binaries in their sample of TTSs could be very low since close binaries lose their disk 
significantly faster than single stars \citep[cf.][]{2006ApJ...653L..57B}. 

We have calculated ages and masses for 241 optically identified individual YSOs classified 
using different schemes (see Table~\ref{sammpleyso}). Here we would like to point out that 
out of six optically identified probable NIR excess stars, five have ages $\lesssim$1 Myr. 
They may be YSOs that are deeply embedded and are formed by the collapse of the core 
of a molecular cloud. Estimated ages and masses of the YSOs range from $\sim$0.1 to 10 Myr 
and $\sim$0.3 to 4.8 M$_\odot$, respectively. This age range indicates a wide spread in the 
formation of stars in the region. The histograms of age and mass distribution of YSOs are 
shown in Fig.~\ref{hist}. 

As stated in Sect.~\ref{distance}, several authors have used different distances 
(2.3 kpc) than ours (2.9 kpc) for the Carina nebula. Therefore, we also examined 
the above results (ages and masses of YSOs) for the distance of 2.3 kpc. The ages and masses 
of YSOs are once again derived using the same procedure as described above and the 
corresponding histograms are overplotted in Fig.~\ref{hist}. As can be seen in 
this figure, both derived values are more or less similar within their corresponding 
errors, although there are slight differences in the number of YSOs that are less than 
1 Myr. By looking at this figure, we can safely conclude that the majority of the YSOs 
are younger than 1 Myr and have a mass lower than 2 M$_\odot$. These age and mass are 
comparable with the lifetime and mass of TTSs.

\subsection{Initial mass function}

The distribution of stellar masses that formed in one star-formation event in a given volume 
of space is called the initial mass function (IMF), and together with the star formation rate, 
the IMF dictates the evolution and fate of galaxies and star clusters. The effects of 
environment may be more revealing at the low-mass end of the IMF, since one might imagine that 
the lower end of the mass spectrum is most strongly affected by external effects. The goals 
of this study are to identify the PMS populations in order to study the IMF down to the 
substellar regime.

The mass function (MF) is often expressed by a power law,
$N (\log m) \propto m^{\Gamma}$ and  the slope of the MF is given as

    $$ \Gamma = d \log N (\log m)/d \log m  $$

\noindent
where $N (\log m)$ represents the number of stars per unit logarithmic mass interval. 

The IMF in the Galaxy has been estimated empirically. The first such determination by 
\citet{1955ApJ...121..161S} gave $\Gamma = -1.35$ for the stars in the mass range 
$0.4 \leq$ M/M$_\odot \leq 10$. However, more recent works \citep[e.g.,][]{1979ApJS...41..513M, 
1986FCPh...11....1S, 1991ARAA..29..129R, 2002Sci...295...82K}  
suggest that the mass distribution deviates from a pure power law. It has been 
shown \citep[see, e.g.,][]{2005ASSL..327.....C, 1986FCPh...11....1S, 1998ASPC..142..201S, 
2002Sci...295...82K, 2003PASP..115..763C} that, for masses above $\sim$1M$_\odot$, the 
IMF can generally be approximated by a declining power law with a slope similar to what 
is found by \citet{1955ApJ...121..161S}. However, it is now clear that this power law does 
not extend to masses much below $\sim$1M$_\odot$. The distribution becomes flatter below 
1 M$_\odot$ and turns off at the lowest stellar masses. 
It has also often been claimed 
that some (very) massive star-forming regions have a truncated IMF, i.e., contain much 
smaller numbers of low-mass stars than expected from the field IMF. However, most of the 
more recent and sensitive studies of massive star-forming regions 
\citep[see, e.g.,][]{2009MNRAS.396.1665L, 2009A&A...501..563E} find the numbers of low-mass 
stars in agreement with the expectation from the ``normal" field star IMF. 
\citet{2011AA...530A..34P} confirm these results for the Carina Nebula and support the 
assumption of a universal IMF (at least in our Galaxy). In consequence, this result also 
supports the notion that OB associations and very massive star clusters are the dominant 
formation sites of the galactic field star population, as already suggested by 
\citet{1978PASP...90..506M}.

\begin{figure}
\centering\includegraphics[height=6.25cm,width=8.5cm,angle=0]{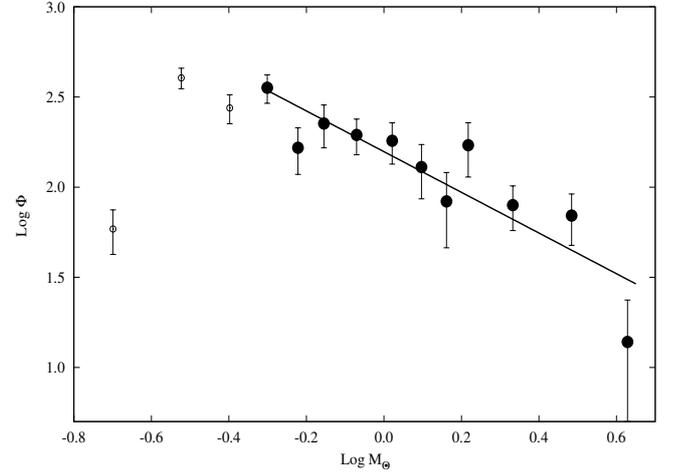}
\caption{\label{mf} Plot of the mass function in the CrW region. Log~$\Phi$ represents
log$N$(log~$m$). The error bars represent $\pm\sqrt N$ errors. The solid line shows a 
least-squares fit over the entire mass range 0.5 $<$ M/M$_\odot <$ 4.8.
Open and filled circles represent the points below and above the completeness limit of our
data, respectively.
}
\end{figure}

We have optically identified 241 YSOs (cf. Sect.~\ref{ysoidf}) in the region of CrW and 
then calculated their masses (cf. Sect.~\ref{ageandmass}) with the help of optical CMD 
using the theoretical PMS of \citet{2000AA...358..593S}. Here we would like 
to mention that for our photometry, the completeness limit is 0.5 M$_\odot$ for a distance 
of 2.9 kpc. The MF of the CrW region is plotted in Fig.~\ref{mf}. The slope of the MF 
`$\Gamma$' in the mass range 0.5 $<$ M/M$_\odot$ $<$ 4.8 comes out to be $-1.13\pm0.20$, 
which is a bit shallower than the value given by \citet{1955ApJ...121..161S}, 
and there seems to be no break in the slope at M $\sim$1 M$_\odot$, as has been noticed in 
previous works \citep{2007MNRAS.380.1141S, 2008MNRAS.383.1241P, 2008MNRAS.384.1675J}. On the 
other hand, \citet{2011AA...530A..34P} show that, down to a mass limit around 0.5 $-$ 1 
M$_\odot$, the shape of the IMF in Carina is consistent with that in Orion (and thus the 
field IMF). Their results directly show that there is clearly no deficit of low-mass stars 
in the CNC down to $\sim$1M$_\odot$.

\subsection{$K$-band luminosity function}\label{kblf}

The $K$-band luminosity function (KLF) is the number of stars as a function of $K$-band 
magnitude. It is frequently used in studies of young clusters and star-forming regions 
as a diagnostic tool of the mass function and the star formation history of their stellar 
populations. The interpretation of KLF has been presented by several authors 
\citep[see, e.g.,][and references therein]{1993prpl.conf..429Z, 2000ApJ...533..358M, 
2003ARAA..41...57L}. 

\begin{figure*}
\centering\includegraphics[height=6cm,width=7cm,angle=0]{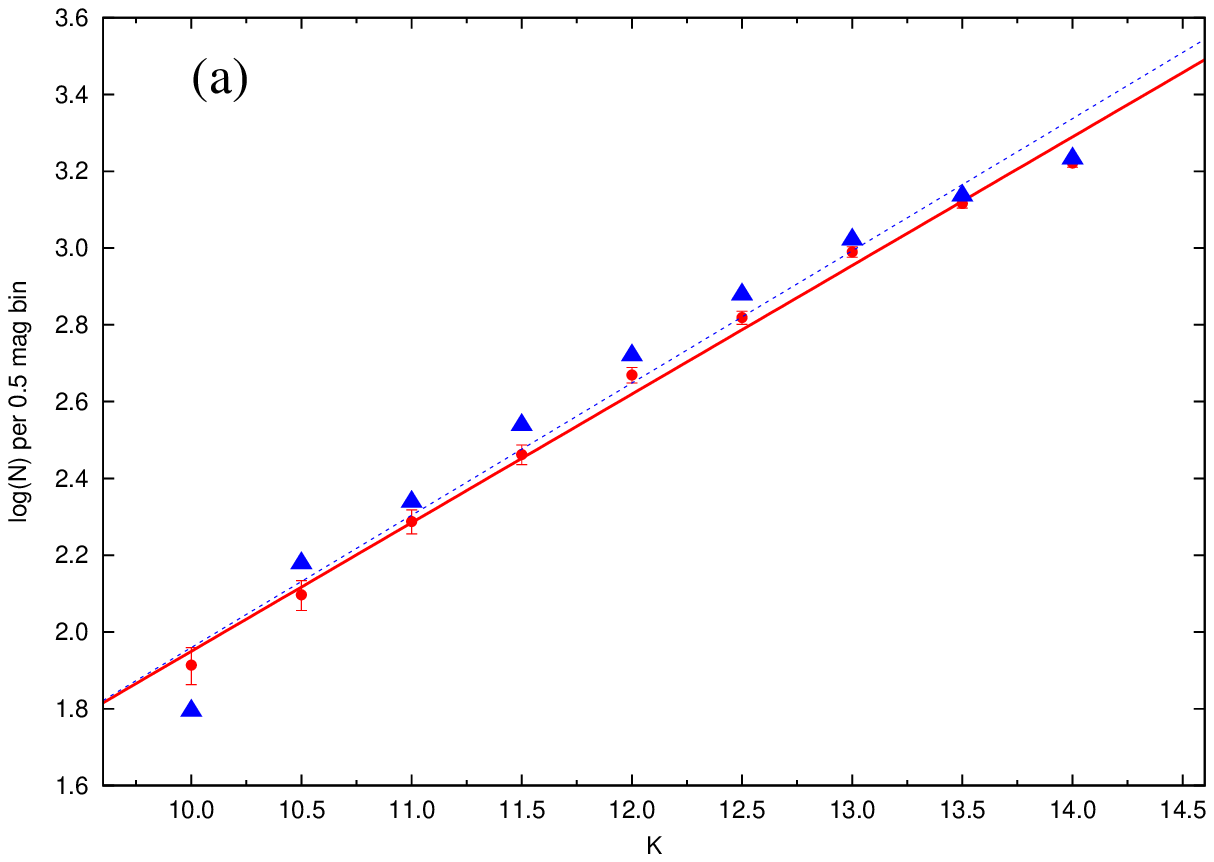}
\centering\includegraphics[height=6cm,width=7cm,angle=0]{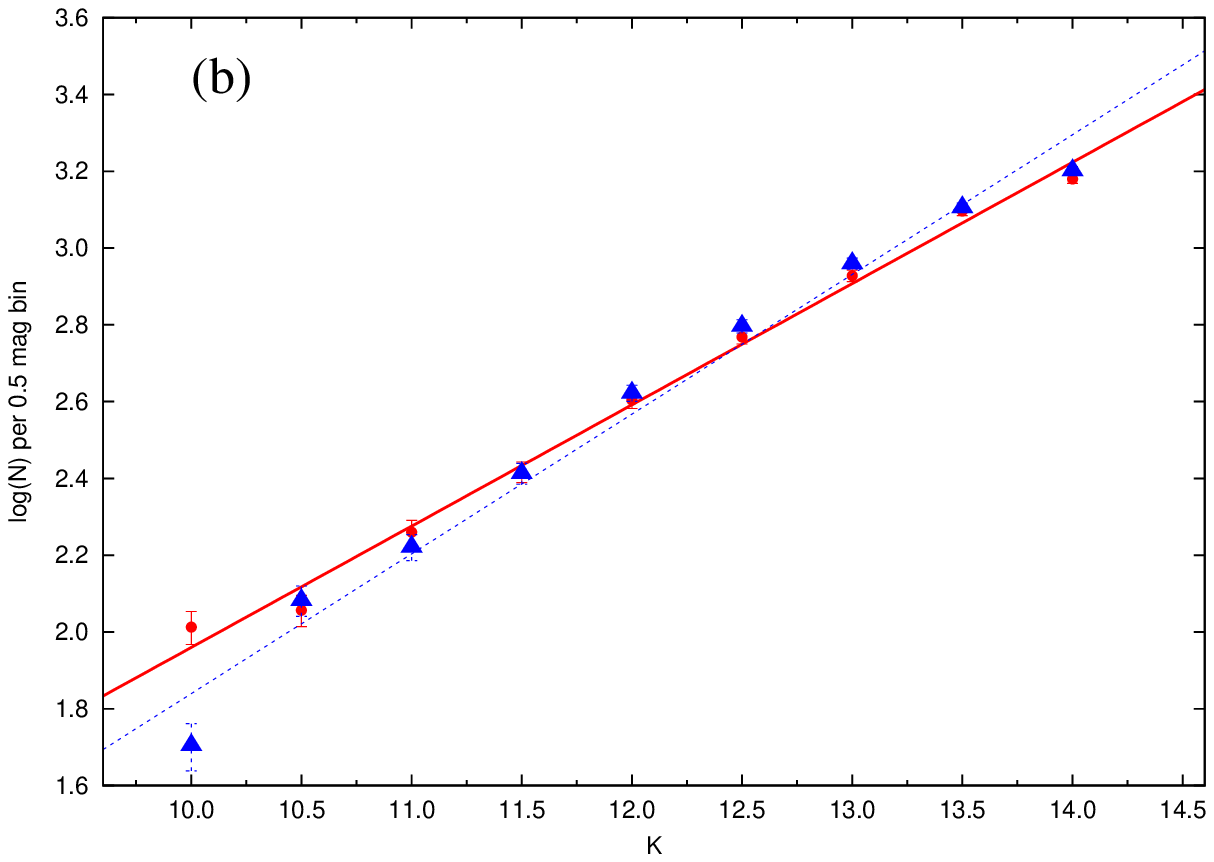}
\caption{\label{klf} (a) Comparison between the observed KLF in the reference field 
(red filled circles) and the simulated KLF from star counts modeling (blue filled triangles). 
If the star counts represent the number N of stars in a bin, the associated error bars are 
$\pm \sqrt N$. The KLF slope ($\alpha$, see Sect.~\ref{kblf}) of the reference field 
(solid line) is $0.34\pm0.01$. The simulated model (dashed line) also gives the same 
value of slope ($0.34\pm0.02$).
(b) The KLF for the CrW region (filled red circles) and the simulated star counts (blue 
filled triangles). In the magnitude range 10.5 $-$ 14.25, the best-fit KLF slope ($\alpha$) 
for the CrW region (solid line) is $0.31\pm0.01$, whereas for the model (dashed line), after
taking extinction into account, it comes out to be $0.36\pm0.02$.}
\end{figure*}

To obtain the KLF, it is essential to take the incompleteness of the data 
and the foreground and background source contaminations into account. The completeness of the data is 
estimated using the {\it ADDSTAR} routine of {\it DAOPHOT} as described in Section~\ref{cod}. To 
consider the foreground/background field star contaminations, we used both the Besan\c{c}on 
Galactic model of stellar population synthesis \citep{2003AA...409..523R} and the nearby reference 
field stars. Star counts are predicted using the Besan\c{c}on model in the direction of the control 
field. We checked the validity of the simulated model by comparing the model KLF with that of 
the control field and found that both KLFs match rather well. An advantage to using the model 
is that we can separate the foreground (d $<$ 2.9 kpc) and background (d $>$ 2.9 kpc) field stars. 
As mentioned in Section~\ref{reddening}, the foreground extinction using optical data was found to 
be $A_V\sim$0.93 mag. The model simulations with $A_V$ = 0.93 mag and d $<$ 2.9 kpc gives the 
foreground contamination.

The background population (d $>$ 2.9 kpc) was simulated with $A_V$ = 3.4 mag 
in the model. We thus determined the fraction of the contaminating stars
(foreground + background) over the total model counts. This fraction was 
used to scale the nearby reference field. The KLF is expressed by the 
power law

${{ \it {d} N(K) } \over {\it{d} K }} \propto 10^{\alpha K}$ ,\\
\noindent \\
where ${ \it {d} N(K) } \over {\it{d} K }$ is the number of stars per 0.5 magnitude
bin, and $\alpha$ is the slope of the power law.

Figures~\ref{klf}a and b show the KLF for the reference field and CrW region, 
respectively. The $\alpha$ for the reference field and simulated model is 
$0.34\pm0.01$ and $0.34\pm0.02$, respectively. Similarly $\alpha$ for the 
CrW region is $0.31\pm0.01$, whereas for the model, after taking the extinction
into account, it comes out to be $0.36\pm0.02$. 

\section{Discussion: star formation scenario in the CrW region}\label{diss}

\citet{2011ApJS..194...14P} using Spitzer MIR data identified 1439 YSOs 
(Pan Carina YSO Catalog) in the field surveyed by the CCCP. The spatial 
distribution of these YSOs throughout the Carina Nebula shows a highly complex 
structure with clustering at several positions. The majority of YSOs identified 
by them are located inside the H{\sc ii} cavities near, but less frequently 
within, the boundaries of dense molecular clouds and the ends of the pillars.  
They also found that the high concentration of the intermediate mass YSOs 
is in Tr~14 itself. They have concluded that the recent star formation history 
in the Carina Nebula has been driven or at least regulated by feedback from 
the massive stars.

\begin{figure*}
\centering\includegraphics[height=14cm,width=16cm,angle=0]{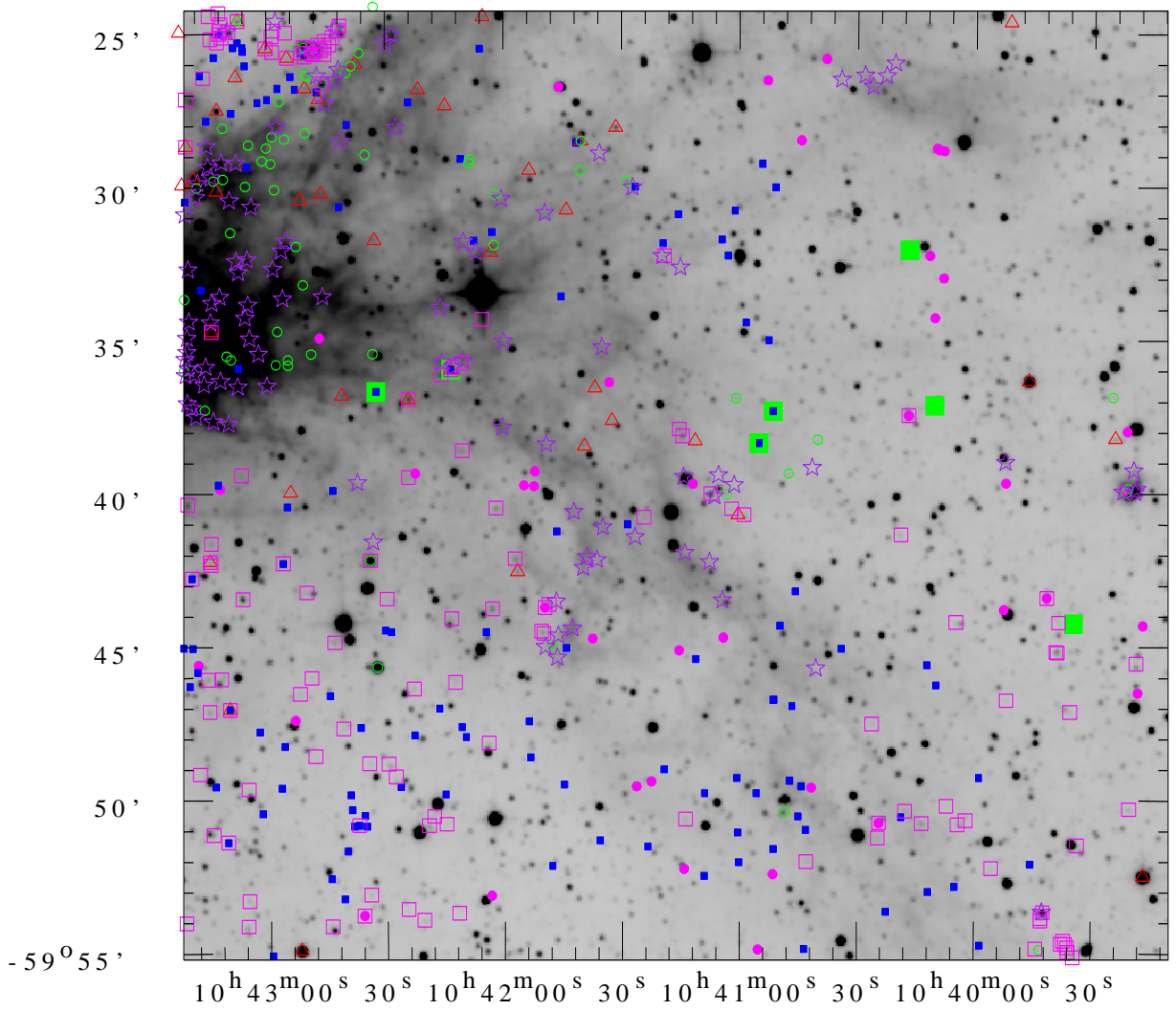}
\caption{\label{spa} Spatial distributions of different classes of YSOs.
Various symbols are overlaid on the {\rm WISE} 4.6 $\mu$m image. The filled square
symbols represent X-ray identified sources ({\it XMM-Newton} bigger green, {\it Chandra}
sources small blue). Open magenta squares, open red triangles, filled magenta 
circles, and open green circles are Spitzer-identified YSOs, CTTSs, $H\alpha$ emission 
stars, and probable NIR-excess YSOs, respectively. Purple star symbols are Herschel 
YSO sources. The abscissae and the ordinates represent RA and Dec, respectively for the
J2000 epoch.}
\end{figure*}

Recently, \citet{2013AA...549A..67G} identified 642 YSOs in the Carina Complex 
with the help of FIR Herschel data. These YSOs are also found to be 
highly heterogeneously distributed in the region, and they do not follow the 
distribution of cloud mass. \citet{2013AA...549A..67G} show that the Herschel 
selected YSO candidates are located near the irradiated surfaces of clouds 
(see Fig.~\ref{spa}) and pillars, whereas the Spitzer selected `YSO' candidates 
\citep{2011ApJS..194...14P} often surround these pillars. This characteristic 
spatial distribution of the young stellar populations in different evolutionary 
stages has been related by 
\citet{2013AA...549A..67G} to the idea that the advancing ionization fronts 
compress the clouds and lead to cloud collapse and star formation in these clouds, 
just ahead of the ionization fronts. They further state that some fraction of the 
cloud mass is transformed into stars (and these are the YSOs detected by Herschel), 
while another fraction of the cloud material is dispersed by the process of 
photo-evaporation. As time proceeds, the pillars shrink, and a population of 
slightly older YSOs is left behind and revealed after the passage of the ionization 
front. Their results provide additional evidence that the formation of these YSOs 
was indeed triggered by the advancing ionization fronts of the massive stars as 
suggested by the theoretical models \citep[see][]{2010ApJ...723..971G}. 

\citet{2013AA...554A...6R} with the help of the wide-field Herschel SPIRE and PACS 
maps, determined the temperatures, surface densities, and the local strength of the 
far-UV irradiation for all the cloud structures over the entire spatial extent of the CNC. 
They find that the density and temperature structure of the clouds in most parts of 
the CNC are dominated by the strong feedback from the numerous massive stars, rather 
than by random turbulence. They also conclude that the CNC is forming stars in 
a particularly efficient way, which is a consequence of triggered star formation by 
radiative cloud compression due to numerous high mass stars.

\begin{figure*}
\centering
\subfigure[a][]{\includegraphics[height=7cm,width=8cm,angle=0]{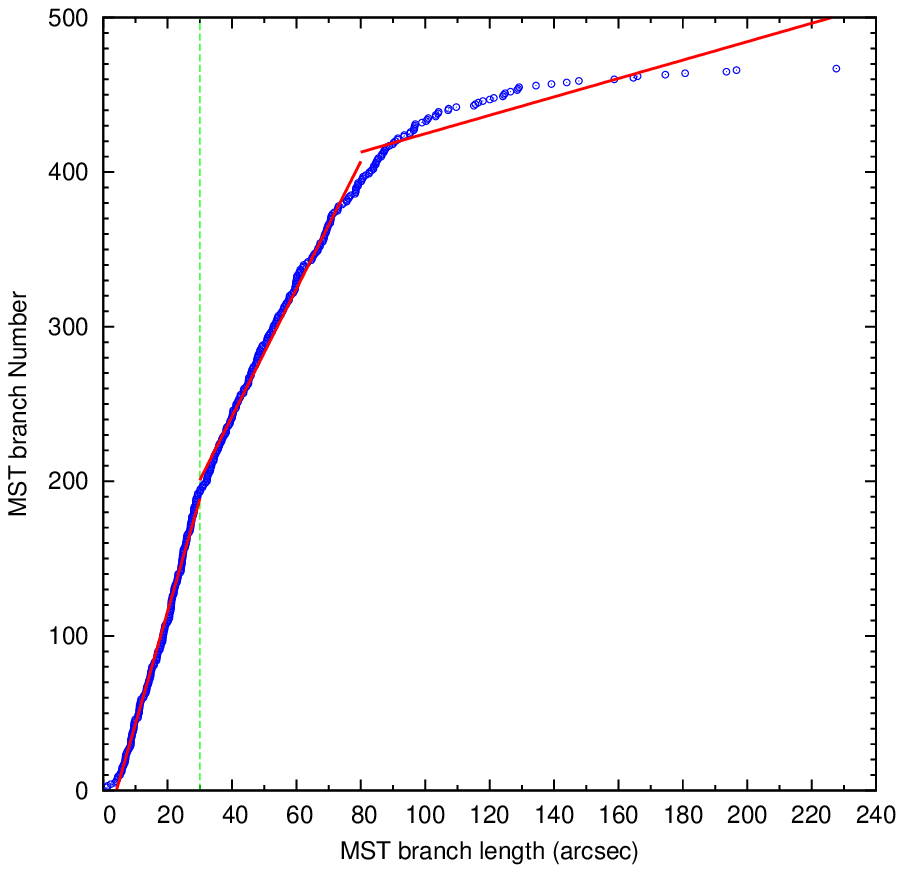}\label{cdf_1}}
\subfigure[b][]{\includegraphics[height=7cm,width=8cm,angle=0]{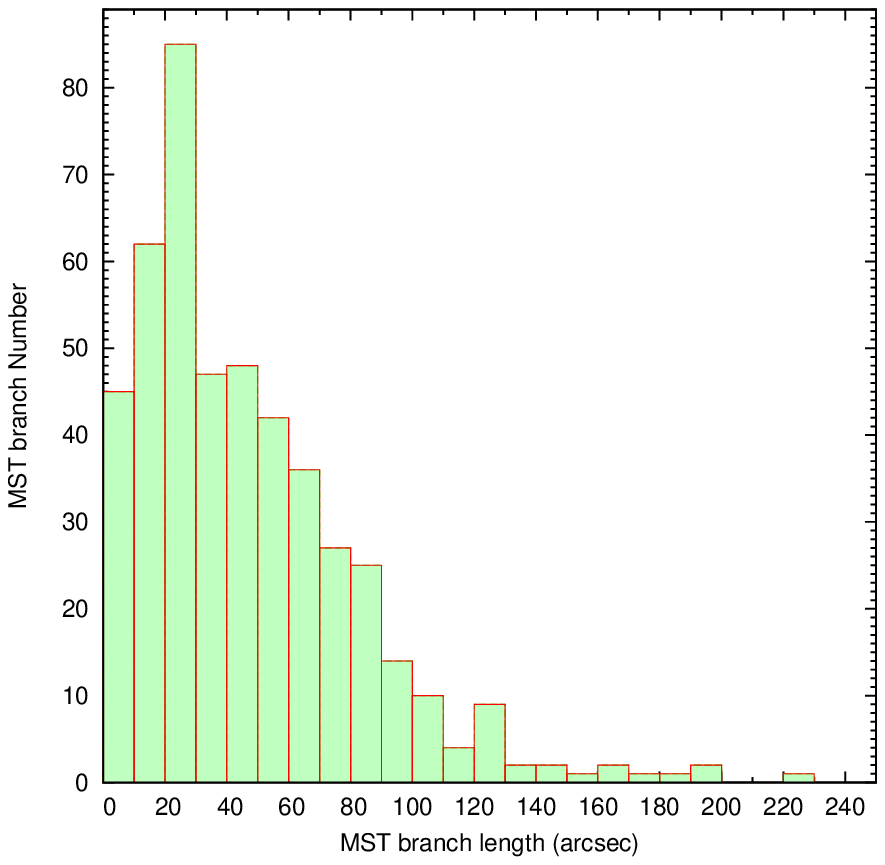}\label{cdf_2}}
\caption{\label{cdf}
Cumulative distribution of the MST branch lengths. In panel (a), the solid lines 
represent the linear fits to the points smaller and larger than the chosen critical branch length. 
The critical radius is shown by a vertical line. Panel (b) is the histogram of the MST 
branch lengths for the YSOs in the CrW region (see text).}
\end{figure*}

In the center of Carina, there are the young clusters, Trumpler 14, 15, and 16, that host 
about 80\% of the high mass stars of the entire complex \citep{2013AA...554A...6R}. 
This is also the hottest region of the nebula with temperatures ranging between 30 and 50 K, 
whereas the molecular cloud at the western side of Tr~14 has a temperature of about 30 K 
and a decrease in density from the inner to the edge part \citep{2013AA...554A...6R}.
Our studied region CrW contains this cloud, which can be seen in our infrared extinction 
map (see Fig.~\ref{av}). In Fig.~\ref{spa}, different classes of YSOs identified in our 
study are overlaid on the {\rm WISE} 4.6 $\mu$m (MIR) image. We can easily see the extension 
of the dust lane in the figure from northeast to southwest of the CrW region. The 
northeast region contains the outer most part of the cluster Tr~14 along with the high 
density region of the molecular cloud (see Fig.~\ref{av}). \citet{2008hsf2.book..138S} 
show the spatial relationship of Tr~14, the ionized gas, the PDR emission, the molecular 
gas, and the dust lane. The brightest molecular emission is concentrated towards the dark 
western dust lane offset from the center of Tr~14 by 4 arcmin. The radio continuum for 
emission source ``Car I" can also be seen here at the interface of the dust lane and 
the bright H{\sc ii} region. Between this source and the molecular cloud, a widespread 
PDR emission can also be seen in the form of an arc like PAH emission feature at 
3.3 $\mu$m \citep{2002MNRAS.331...85R}. At a projected distance of $\sim$2 pc, the UV 
output of Tr~14 dominates the other Carina Nebula clusters such as Tr~16 in 
determining the local flux at the PDR in the northern cloud 
\citep{2003AA...412..751B, 2006MNRAS.367..763S, 2008hsf2.book..138S}.
This spatial sequence of Tr~14, radio source, PAH emission, and then strong molecular 
emission delineates a classical edge-on PDR \citep{2003AA...412..751B}. The edge of 
this region contains many Spitzer-identified YSOs. The alignment of the YSOs in this 
region may be due to the star formation triggered by high mass stars of Tr~14. 

The Herschel-identified YSOs \citep{2013AA...549A..67G} are located mainly in the high 
density region of the molecular clumps and in small groupings at several places along the 
dust lane. \citet{2013AA...549A..67G} have derived an age of $\sim$0.1 Myr for their sample of
YSOs. The probable NIR excess stars identified in this study also follow this region.
For some of them, we derived ages $\lesssim$1 Myr. These sets of identified YSOs are 
basically very young in nature and are embedded in the cores of the  molecular cloud. 
We could not say anything about the northwest region of CrW, which is not well covered 
by previous surveys.

\citet{2010MNRAS.406..952S} observed in their 
western mosaic (which contains most of our observed region including the dark lane, 
see their Fig. 3), the YSOs density of around 500 sources/deg$^2$ with little signs of 
clustering. In this study we have identified 467 YSOs falling in the CrW region.
The overall density of this region then turns out to be $\sim$1700 sources/deg$^2$ which 
is higher than three times the YSOs density given by \citet{2010MNRAS.406..952S}. Here it 
is worthwhile to mention that the PCYC used in previous studies has a sensitivity 
problem at the ionization front between Tr~14 and the Car I molecular cloud core to 
the west \citep{2005ApJ...634..476Y, 2007AA...476..199A} where the diffuse MIR nebular
emission is bright \citep{2011ApJS..194...14P}.  

\begin{table*}
\centering
\caption{The YSO cores identified in the CrW region and their characteristics.}
\label{ccore}
\begin{tabular}{ccccc}
\hline \hline
Core    & Radius  & Number of YSOs & Number            &Median             \\
number  & (pc)    &  in core (N)   & density (N/pc$^2$)&branch length (pc) \\  \hline
A       &  0.45   &  6             &  9.43             &  0.32  \\
B       &  0.37   &  6             &  13.95            &  0.26  \\
C       &  0.51   &  8             &  9.79             &  0.36  \\
D       &  0.56   &  7             &  7.11             &  0.33  \\
E       &  0.18   &  4             &  39.30            &  0.12  \\
F       &  0.26   &  5             &  23.54            &  0.20  \\
G       &  0.39   &  4             &  8.37             &  0.36  \\
H       &  0.49   &  6             &  7.95             &  0.31  \\
I       &  0.64   &  6             &  4.66             &  0.30  \\
J       &  0.44   &  7             &  11.51            &  0.28  \\ \hline
Average &  0.43   &  5.9           &  13.56            &  0.28  \\
\hline
\end{tabular}
\end{table*}

The complex observational patterns (e.g., filaments, bubbles, and irregular clumps, etc.) 
in a molecular cloud such as Carina nebula are resulting from the interplay of fragmentation 
processes. The star formation usually takes place inside the dense cores of the molecular 
clouds, and the YSOs often follow clumpy structures of their parent molecular clouds 
\citep[see, e.g.,][]{1993AJ....105.1927G, 1996AJ....111.1964L, 1998A&A...336..150M,
2002ApJ...566..993A, 2005ApJ...632..397G, 2006ApJ...636L..45T, 2007ApJ...669..493W, 
2008ApJ...674..336G}.
Recently, fragmentations in gas with turbulence \citep[e.g.,][]{2007prpl.conf...63B} and 
magnetic fields \citep[e.g.,][]{2007prpl.conf...33W} have been discussed, leading to detailed
predictions of the distributions of fragment spacings. The spatial distribution of YSOs in 
a region can be analyzed in terms of a typical spacing between them in order to compare this 
spacing to the Jeans fragmentation scale for a self-gravitating medium with thermal pressure 
\citep{1993AJ....105.1927G}. Some recent observations of star-forming regions have been
analyzed in terms of the distribution of nearest neighbor (NN) distances 
\citep[see][]{2005ApJ...632..397G, 2006ApJ...636L..45T} and find a strong peak in their 
histogram of NN spacings for the protostars in young embedded clusters. This peak indicated 
a significant degree of Jeans fragmentation, since this most frequent spacing agreed with an 
estimate of the Jeans length for the dense gas within which the YSOs are embedded. These 
results also suggest that the tendency for a narrow range of spacings among YSOs in a cluster 
can last into the Class II phase of YSO evolution.

\begin{figure*}
\centering
\includegraphics[height=16cm,width=16cm,angle=0]{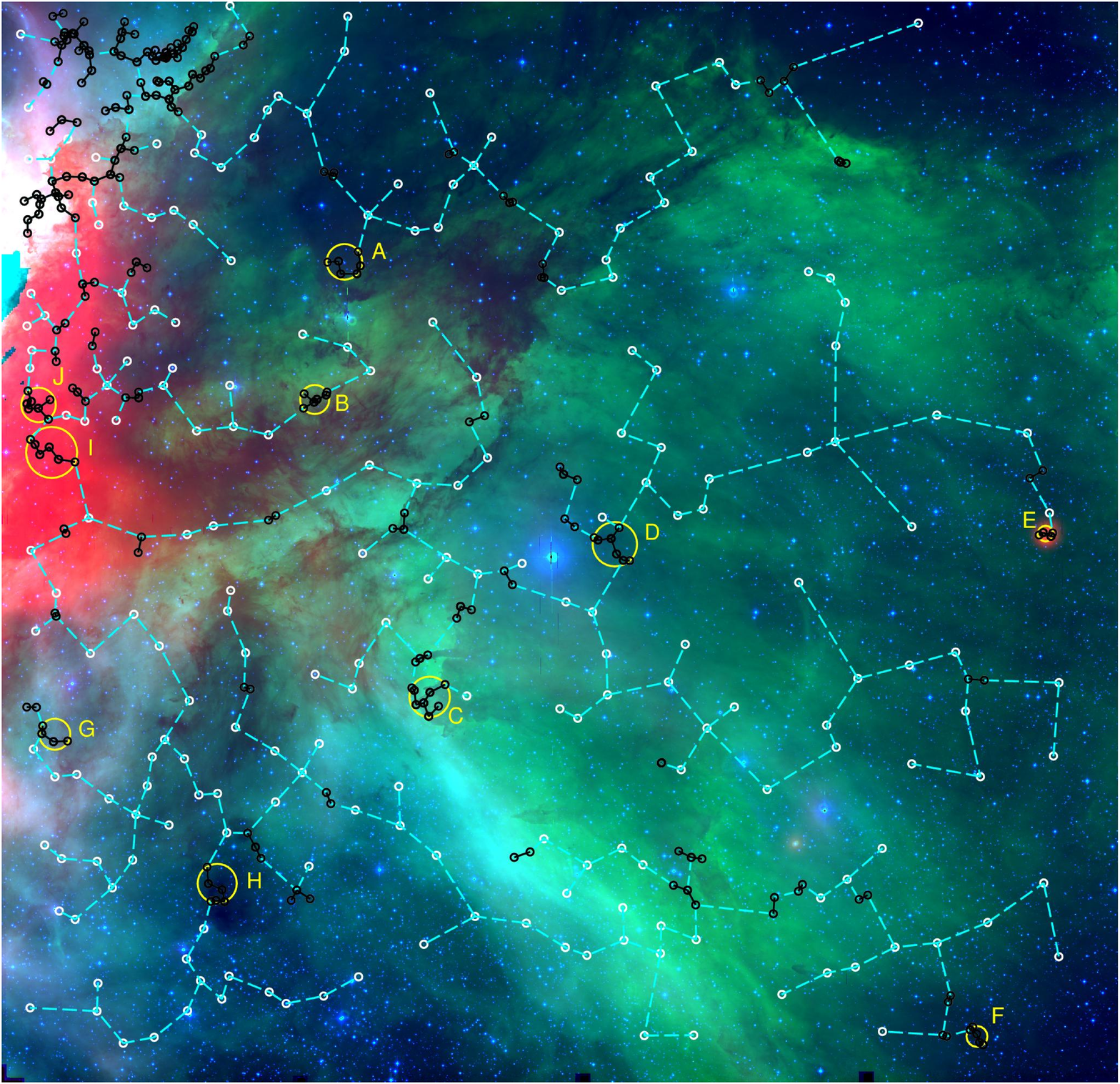}
\includegraphics[height=7.6cm,width=9.2cm,angle=0]{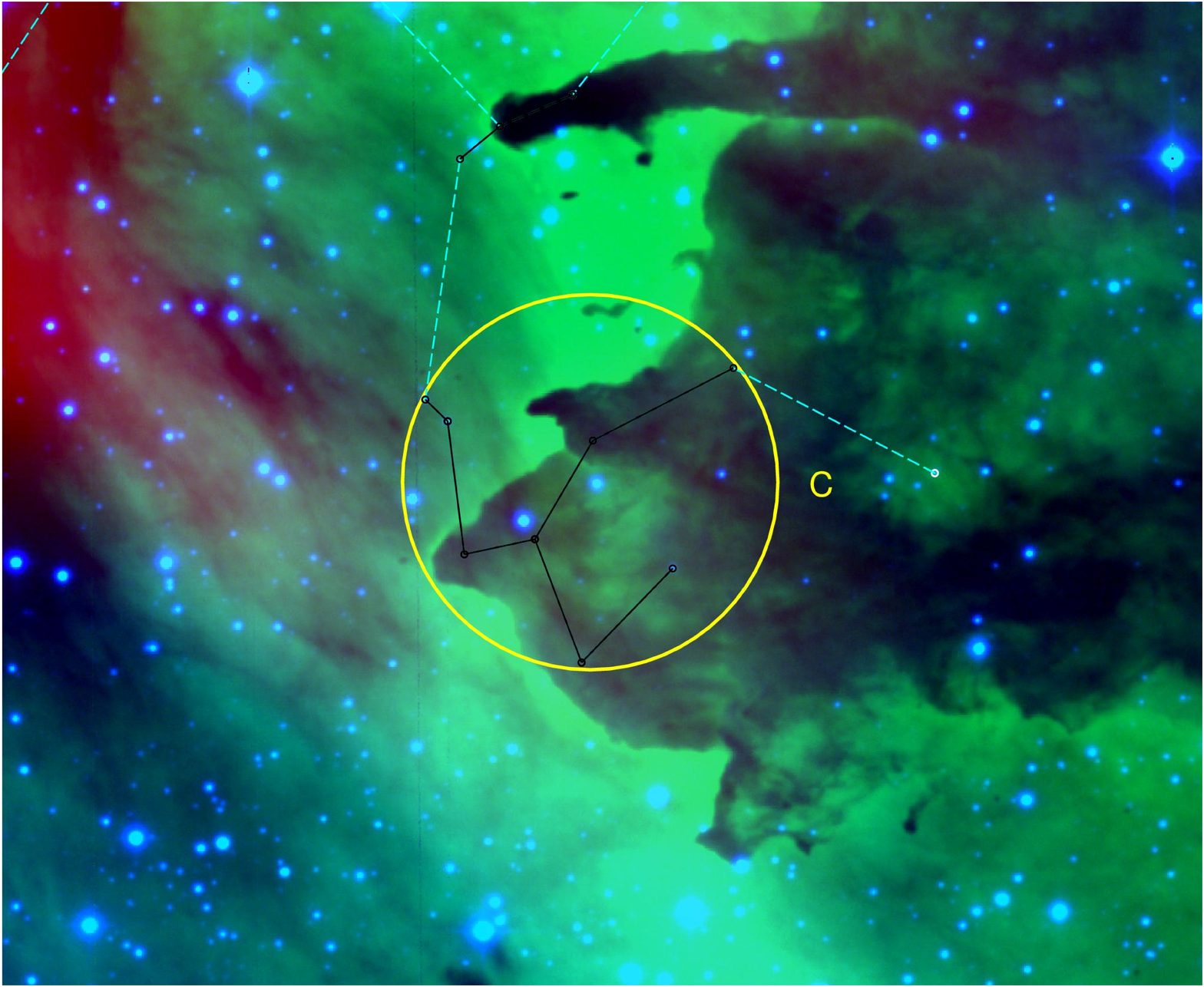}
\includegraphics[height=7.6cm,width=8.5cm,angle=0]{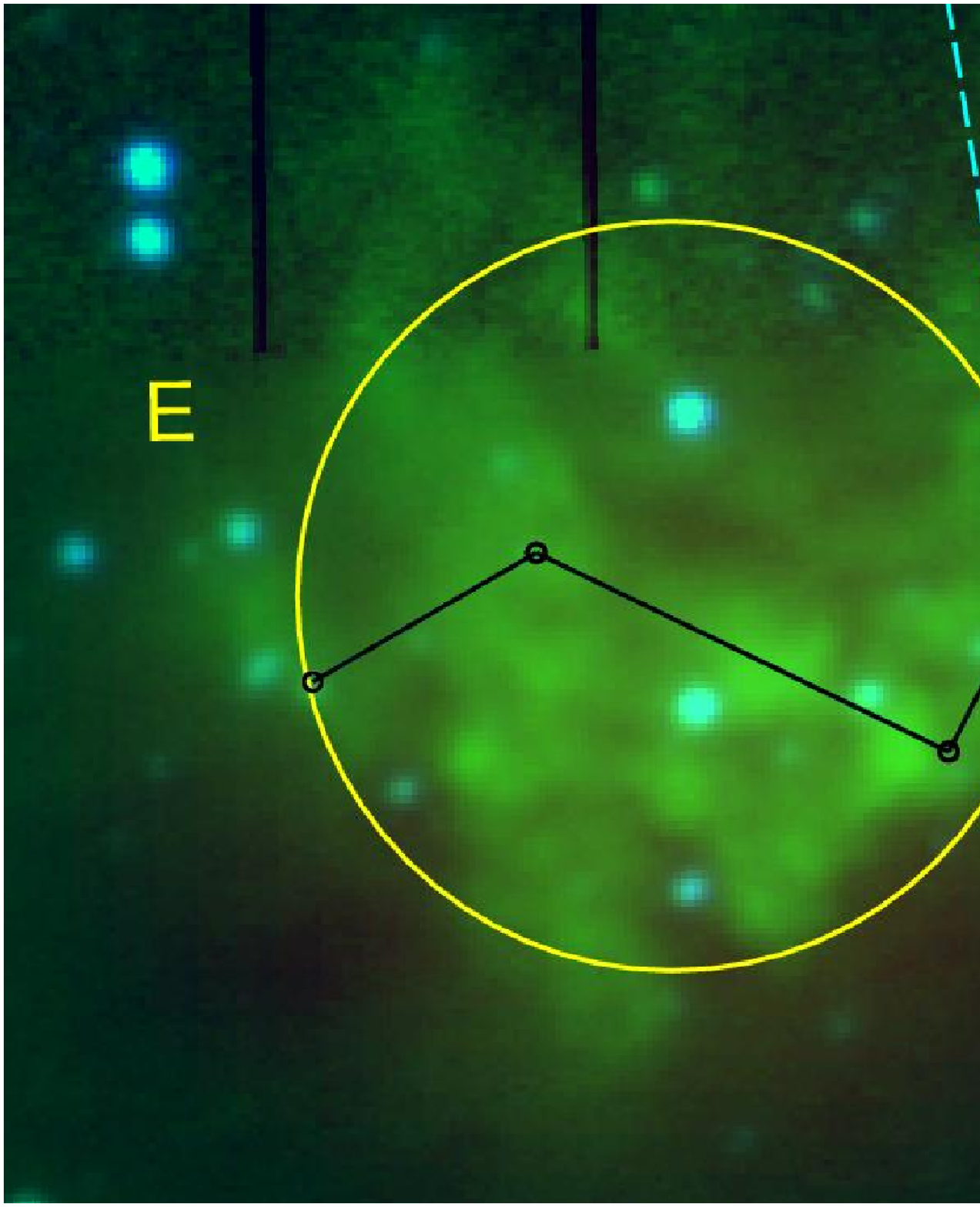}
\caption{\label{mst}
Top: Minimal spanning tree of the YSOs overplotted on a color composite image of the CrW 
region ({\rm WISE} 22 $\mu$m (red), $H\alpha$ band (green), and $V$ band
(blue) images). WR~22 is situated in the center. The white circles connected with dotted lines, 
and black circles connected with solid lines are the branches that are larger and smaller than 
the basic critical length, respectively. The identified ten cluster cores are encircled with 
yellow color and labeled with A to J. Bottom: Two zoomed images of YSO cores, C and E, are 
shown in the lower left and right panels, respectively (see text for detail).}
\end{figure*}

Recently, \citet{2009ApJS..184...18G} have done a complete characterization of 
the spectrum of source spacings using the minimal spanning tree (MST) of source positions. 
The MST is defined as the network of lines, or branches, that connect a set of points together 
such that the total length of the branches is minimized and there are no closed loops 
\citep[see, e.g.,][and references therein]{2004MNRAS.348..589C, 2009ApJS..184...18G}.
\citet{2009ApJS..184...18G} demonstrate that the MST method yields a more 
complete characterization than NN method. Therefore, for the present study, we used the 
same MST algorithm to analyze the spatial distribution of YSOs in the CrW region.

In Fig.~\ref{cdf}b, we plotted the histogram of MST branch lengths for the YSOs in the 
CrW region. From this plot, it is clear that they have a peak at small spacings and that 
they also have a relatively long tail of large spacings. Peaked distance distributions 
typically suggest a significant subregion (or subregions) of relatively uniform, elevated
surface density. By adopting an MST length threshold, we can isolate those sources that are closer
together than this threshold, yielding populations of sources that make up a local surface density
enhancement. To get this threshold distance, in Fig.~\ref{cdf}a, we plotted the cumulative 
distribution function (CDF) for the branch length of YSOs. The curve shows three different slopes: 
first a steep-sloped segment at short spacings then a transition segment that approximates the 
curved character of the intermediate-length spacings, and finally a shallow-sloped segment at long 
spacings. For the present study, we have chosen the peak in the histogram ($\sim$30 arcsec 
$\sim$0.42 pc), which corresponds to the first steep-sloped segment, as a threshold or critical 
length.

In Fig ~\ref{mst}, the results from MST analysis are overplotted on the RGB image of the 
CrW region. This image was created using red, green, and blue colors for the {\rm WISE} 22 $\mu$m, 
$H\alpha$, and $V$ band images, respectively. Circles and MST connections in black deal with 
the objects that are more closely spaced than the critical length (0.42 pc). 

A close inspection of Fig.~\ref{mst} reveals that the CrW region exhibits heterogeneous structures, 
and there are several close concentrations of YSOs distributed along the molecular clumps. Prominent
YSO clustering can be seen in the northeastern part of this figure and are possibly part of the star 
cluster Tr~14. There are also ten cores (having MST branch lengths less than the critical distance) 
distributed along the molecular cloud (indicated in Fig.\ref{mst}). 
A close view of the cores C and E  can be seen in the lower left and right panels in Fig~\ref{mst}.
The details about these cores have been given in Table \ref{ccore}. The majority of the 
members of all these cores are the YSOs identified in the Herschel survey having very 
young ages. Our result agrees with the conclusions of \citet{2009ApJS..184...18G} and 
\citet{2012AJ....144..101G}, indicating that the young protostars are found in a region having 
marginally higher surface densities than the more evolved PMS stars. The average of median branch 
length and core radius is found to be 0.28 pc and 0.43 pc, respectively (see Table \ref{ccore}).

\begin{figure*}
\centering\includegraphics[height=14cm,width=16cm,angle=0]{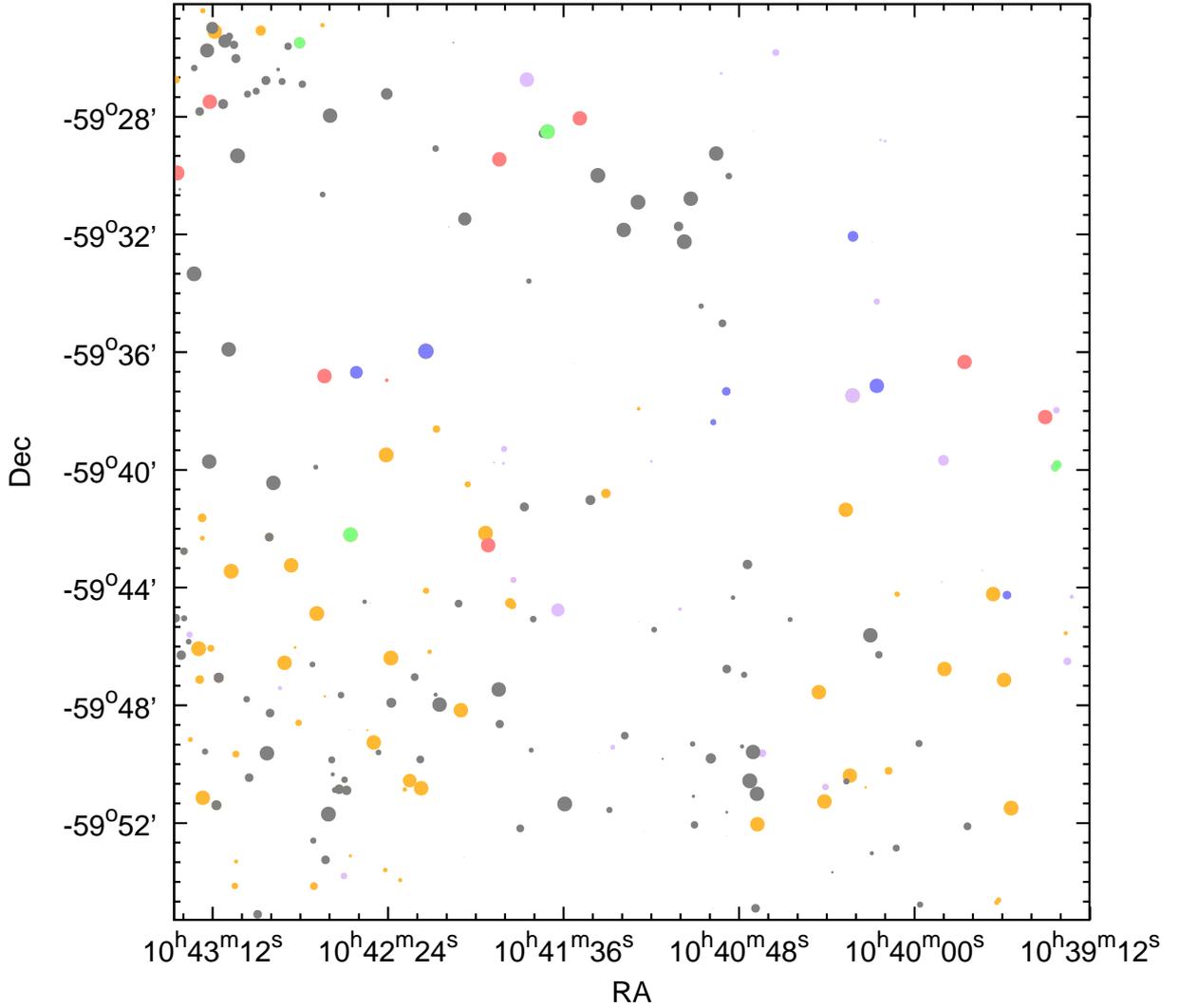}
\caption{\label{age} Spatial distribution of the optically identified YSOs in the CrW region.
The size of the symbols represents the age of the YSO, i.e. bigger the size younger the YSO is.
Various colors represent YSOs identified using different schemes (Spitzer - orange,
$H\alpha$ - purple, CTTS - red, {\it Chandra} sources - black, {\it XMM-Newton} - blue, and 
IR excess - green).}
\end{figure*}


To check the role of WR~22 in the formation of stars in this region, we tried to look 
at the spatial distribution of YSOs (Fig.~\ref{age}) with optical counterparts (whose age has 
been derived using the $V/(V-I)$ CMD, cf. Sect.~\ref{optcmd}) in the CrW region. The general 
observation is that within the detection limit of optical observations, the YSOs show 
mixed populations of different ages throughout the CrW region.
The extreme northeastern region contains a group of YSOs that are under the direct influence 
of the high-mass stars of Tr~14, and it also shows a mixed population. For most of the YSOs in 
the dust lane, we have not found their optical counterparts. This age distribution around CrW 
does not show any trend even in the presence of a very massive star such as WR~22. It seems 
that WR~22 has not influenced these YSOs much. 
\citet{2010MNRAS.406..952S} presented Spitzer observations of a part of the CrW region studied here. 
These authors reached similar conclusions, suggesting that this very massive star may be projected in 
the foreground or background compared to the surrounding molecular gas, or it could have only 
recently arrived at its present location.

\section{Summary and conclusions }

Although the center of the Carina nebula has been studied extensively, the outer region 
has been neglected due to the absence of wide field optical surveys. In this study, we 
investigated a wide field ($32' \times 31'$) located in the west of the Carina nebula 
and centered on the massive binary WR~22. 
To our knowledge, this is the first detailed study of this region. We used deep 
optical ($UBVRI$) and $H\alpha$ photometric data obtained with the WFI instrument at the 
ESO/MPG 2.2 m telescope (La Silla). Our $V$ band photometry is complete up to $\sim$21.5 
mag. Low-resolution spectroscopy along with {\it Chandra}, {\it XMM-Newton}, and 2MASS archival 
data sets, was also used in this analysis. We generated various combinations of optical 
and NIR TCD, CMDs and calculated several parameters such as reddening, reddening law, etc. We 
also identified the YSOs located in the region and studied their spatial distribution 
using the MST method. Ages and masses of the 241 YSOs having optical counterparts,
were derived based on $V/(V-I)$ CMD. These YSOs have been further used to constrain the IMF of 
the region. The main scientific results from our study are as follows

\begin{itemize}

\item
The region shows a large amount of differential reddening with minimum and maximum 
$E(B-V)$ values of 0.25 and 1.1 mag., respectively. This region shows an unusual 
reddening law with a total-to-selective extinction ratio $R_V = 3.7\pm 0.1$. 

\item
The MK spectral types for a subsample of 15 X-ray emitting sources in the CrW 
region are established that indicates that the majority of them are late spectral type stars. 
There are three sources belonging to each O, A, and F spectral types; however, six sources are 
of spectral type G.

\item
We cross-correlated the 43 {\it XMM-Newton} X-ray sources from \citet{Claeskens2011} 
with our optical photometry and found that 34 of them are well matched. Out of these 
34 sources, 7 have been identified as YSOs. We also cross-identified the {\it Chandra} X-ray
sources (1465 in our region) with our source list and found 469 objects with optical/NIR 
counterparts. In total, 119 X-ray sources are identified as YSOs, four of them in common with 
{\it XMM-Newton}.

\item
We collected a sample of 467 YSOs identified in the CrW region based on their 
IR-excess, $H\alpha$ and X-ray emission. In some cases, the same YSOs were
identified in more than one scheme. Out of these, there are 41 $H\alpha$ emitters, 
105 Herschel identified YSOs, 136 Spitzer identified YSOs, and 225 2MASS identified 
YSOs in our list. The YSO density for the CrW region turns out to be $\sim$1700 
sources/degree$^2$, which is higher when compared to the reported values 
(500 sources/degree$^2$, \citealt{2010MNRAS.406..952S}).

\item
We calculated the age and mass of 241 individual optically identified YSOs. 
Estimated ages and masses of the YSOs range from $\sim$0.1 to 10 Myr and $\sim$0.3 to 
4.8 M$_\odot$, respectively. This age range indicates a wide spread in the formation 
of stars in the region. The majority of these YSOs are younger than 1 Myr, and their 
mass is below 2 M$_\odot$. 

\item
We derived the IMF and calculated the slope `$\Gamma$' in the CrW region. In the mass 
range $0.5 <$ M/M$_\odot < 4.8$, it comes out as $-1.13\pm0.20$ which is a bit 
shallower than the value of $-1.35$ given by \citet{1955ApJ...121..161S}, and there seems 
to be no break in the slope at M $\sim$1M$_\odot$. The slope of the K-band luminosity 
function is found to be $\alpha = 0.31\pm0.01$.

\item
The spatial distribution of all 467 YSOs has been studied in detail. 
The edge of the irradiated surface between Tr~14 and the molecular cloud contains 
many Spitzer identified YSOs whose formation was probably triggered by the 
high-mass stars of Tr~14. 
The high-density region of molecular clumps contains many probable NIR excess 
stars, as well as Herschel-identified YSOs that are very young in age ($\lesssim$1 Myr).

\item
We used the well-established MST method to identify local density enhancements 
in the YSO distributions. The northeastern part of the studied region presents a more 
prominent YSO clustering. However, there are at least ten cores of four or more very 
young YSO members distributed all over the CrW region and having different core radii. 
The average core radii and median branch length values for these cores are found to 
be 0.43 pc and 0.28 pc, respectively.
The YSOs having optical counterparts in CrW are uniformly distributed having mixed 
population of different ages. The age distribution around CrW does not show any trend 
in the presence of the very massive star WR~22. It seems that WR~22 is a 
foreground/background star which has not influenced the formation of YSOs in the CrW 
region.

\end{itemize}

\section*{Acknowledgements}
The authors are grateful to Mauricio Tapia for sharing his data. 
We thank the anonymous referee for his/her critical review and constructive 
suggestions that helped to improve the content and presentation of the paper.
The observations reported in this paper were obtained with the MPG/ESO 2.2 m Telescope, 
the New Technology Telescope (NTT), and the Very Large Telescope array (VLT) at 
the European Southern Observatory (ESO, Chile). This publication makes use of data 
products from the Two Micron All Sky Survey, which is a joint project of the University of 
Massachusetts and the Infrared Processing and Analysis Center/California Institute of 
Technology, funded by the National Aeronautics and Space Administration and the National 
Science Foundation. The authors (BK, JM, EG, GR, and YN) acknowledge the support of an 
Action de Recherche Concert\'ee (Acad\'emie Wallonie-Europe). JM, EG, GR, and YN
also acknowledge the FNRS and PRODEX contracts.

\bibliographystyle{aa}
\bibliography{wr22}
\end{document}